\documentclass[11pt,a4paper]{article}
\pdfoutput=1

\usepackage{jcappub}
\usepackage[margin=0.95in]{geometry}

\usepackage[utf8]{inputenc}
\usepackage{graphicx}
\usepackage{verbatim}
\usepackage{epsfig}
\usepackage{subfigure}
\usepackage{color}
\usepackage{multirow}
\usepackage{comment}
\usepackage{slashed}
\usepackage{amsmath}
\usepackage{ulem}

\usepackage{amssymb,amsmath,amsfonts,mathrsfs}
\usepackage[dvipsnames]{xcolor}
\usepackage{graphicx}
\usepackage{longtable}
\usepackage{verbatim}
\usepackage{color}
\usepackage[mathscr]{euscript}
\usepackage{framed}
\usepackage{mdframed}  

\usepackage{cancel}
\usepackage{color}
\usepackage{hyperref}
\hypersetup{colorlinks, citecolor=bluscuro, linkcolor=black, urlcolor=bluscuro}
\definecolor{rossos}{cmyk}{0,1,1,0.55}
\definecolor{bluscuro}{rgb}{0.15, 0.2, .85}
\definecolor{bluchiaro}{cmyk}{1,.3,0.,0.1}
\graphicspath{{./Figures/}}

\numberwithin{equation}{section}
\renewcommand\theequation{\arabic{section}.\arabic{equation}}

\usepackage{tikz}

\usepackage{tcolorbox}

\newcommand{\nn}{\nonumber}
\DeclareMathOperator{\re}{Re}
\newcommand{\I}[0]{\mathcal{I}}
\newcommand{\lp }[0]{\left (}
\newcommand{\rp }[0]{\right )}

\newcommand{\dd}{{\rm d}}

\newcommand{\vp}{\vec{p}}
\newcommand{\vk}{\vec{k}}
\newcommand{\vx}{\vec{x}}
\newcommand{\vy}{\vec{y}}

\newcommand{\be}{\begin{equation}\begin{aligned}}
\newcommand{\ee}{\end{aligned}\end{equation}}

\newcommand{\bbe}{\begin{align}}
\newcommand{\eee}{\end{align}}

\newcommand{\bea}{\begin{eqnarray}}
\newcommand{\eea}{\end{eqnarray}}

\def\beq{\begin{equation}}
\def\eeq{\end{equation}}

\def\vp{\varphi}

\def\d{{\rm d}}

\def\vk{{\vec{k}}}

\renewcommand{\k}{\vec{k}}

\def\beqa{\begin{eqnarray}}

	\def\eeqa{\end{eqnarray}}
	
\def\lsim{\mathrel{\rlap{\lower4pt\hbox{\hskip0.5pt$\sim$}}
		\raise1pt\hbox{$<$}}}         
\def\gsim{\mathrel{\rlap{\lower4pt\hbox{\hskip0.5pt$\sim$}}
		\raise1pt\hbox{$>$}}}         

\def\vp{\varphi}

\def\d{{\rm d}}
\def\vx{{\vec{x}}}
\def\vy{{\vec{y}}}

\def\vk{{\vec{k}}}

\def\vp{{\vec{p}}}

\def\d{{\rm d}}

\def\CB{{\cal B}}

\def\eeqa{\end{eqnarray}}

\def\bq{\begin{quote}}
\def\eq{\end{quote}}

 at 10truept
\newcommand{\arXiv}[2]{\href{http://arxiv.org/pdf/#1}{{\tt [#2/#1]}}}
\newcommand{\arXivold}[1]{\href{http://arxiv.org/pdf/#1}{{\tt [#1]}}}

\title{ \huge Cosmological Shapes of Higher-Spin Gravity}

\author[a]{D. Anninos,}
\author[b]{V. De Luca,}
\author[b]{G. Franciolini,}
\author[c]{A. Kehagias,}
\author[b]{A. Riotto}

\affiliation[a]{Department of Mathematics, King's College London, Strand, London WC2R 2LS, UK}
\affiliation[b]{
	Department of Theoretical Physics and Center for Astroparticle Physics (CAP) \\
			24 quai E. Ansermet, CH-1211 Geneva 4, Switzerland}
\affiliation[c]{Physics Division, National Technical University of Athens 15780 Zografou Campus, Athens, Greece}

\abstract{
We explore non-Gaussian features of a massless spin-two field in the Vasiliev theory of higher-spin gravity. The theory contains an infinite tower of interacting gauge fields with increasing spin, and admits four-dimensional asymptotically de Sitter configurations. Using a recent proposal for calculating late-time quantum correlations in Vasiliev theory, we provide an exact formula for the tensor non-Gaussianities of the massless spin-two graviton field. 
By general symmetry considerations, we relate our result to that produced by a tree-level calculation in a gravitational theory containing an Einstein term and a term cubic in the Weyl tensor. The relative coefficient between the two terms is calculated explicitly, exhibiting a significant contribution from the Weyl cubed term. We discuss potential cosmological implications of our results.
}

\emailAdd{dionysios.anninos@kcl.ac.uk}
\emailAdd{valerio.deluca@unige.ch}
\emailAdd{gabriele.franciolini@unige.ch}
\emailAdd{kehagias@central.ntua.gr}
\emailAdd{antonio.riotto@unige.ch}

\begin{document}

\maketitle
\flushbottom

\newpage

\section{Introduction}
\label{sec:intro}
Higher-spin theories are field theories containing an infinite tower of interacting gauge fields with increasing spin. Their spectrum includes a massless spin-two field.
In Minkowski space, one can devise no-go theorems \cite{w,Coleman:1967ad,Weinberg:1980kq} which render the S-matrix of a theory with higher-spin conserved charges essentially trivial. 
These no-go theorems can be avoided  if the spacetime is no longer asymptotically flat \cite{Bekaert:2010hw}. Indeed, non-linear classical equations of motion for interacting massless higher-spin fields in a curved background were obtained by Vasiliev \cite{v1,v2,v3}. 
Here, we consider those Vasiliev theories which admit a four-dimensional de Sitter solution \cite{Vasiliev:1986td,Iazeolla:2007wt}.

That Vasiliev's higher-spin theory admits asymptotically de Sitter configurations may be of interest for cosmology. There are important observational indications that our universe underwent a period of accelerated expansion -- the primordial inflationary era -- at an early stage. During this period, the universe is described by an approximately de Sitter spacetime (for a review see \cite{lr}).  Consequently, we are prompted to ask whether the inflationary era could be of the higher-spin type, and if so, what would the observational imprints for such a scenario be \cite{chen,col,am}.

In this paper we take a few steps toward addressing this question. Concretely, we calculate and analyse the exact three-point cosmological correlator of the higher-spin graviton in the minimal Vasiliev theory whose spectrum comprises a particle for every even spin.\footnote{For inflationary signatures of massive or partially massless higher-spin fields,  see  \cite{am,h1,h2,h22,h3,h4,h5,h6,h7,h8,h9,h10,h11}.} In order to analyse the graviton non-Gaussianities, we use the framework of \cite{anninos} which we occasionally refer to as the $Q$-formalism.  
The $Q$-formalism circumvents the formidable task of calculating quantities in the original Vasiliev formalism \cite{Giombi:2009wh} by exploiting the description of higher-spin gravity theory in terms of a vector-like conformally invariant theory.  
The basic proposal of \cite{anninos} is that the microscopic operator content of the higher-spin theory is given by $2N$ local Hermitian operators $\hat{Q}^\alpha(\vec{x})$, and other $2N$ local Hermitian operators $\hat{\Pi}^\alpha(\vec{x})$ satisfying a non-trivial operator algebra. Higher-spin fields are identified with specific composite operators built from the $\hat{Q}$ and $\hat{\Pi}$. 
One further identifies a particular quantum state in the Hilbert space acted on by the $\hat{Q}$ and $\hat{\Pi}$ with the de Sitter invariant Bunch-Davies state, which turns out to be a Gaussian functional in $Q$. In doing so, the $Q$-formalism provides a systematic framework to extract exact expressions for late-time cosmological correlation functions. The results obtained through the $Q$-formalism agree with all previously known tree-level calculations \cite{stromanninos,Anninos:2012ft,Anninos:2013rza,Bobev:2016sap}. Moreover, the correlators calculated by the $Q$-formalism
 encode non-linear effects of the higher-spin theory, such as those produced by the exchange of an infinite tower of massless higher-spin fields. 

Our results are nicely complemented by a general symmetry argument. It was shown in \cite{mp} that the de Sitter isometries fix, perturbatively, the possible shapes of the three-point graviton correlators to two possible structures. These can be obtained from a tree-level calculation in the following theory%
\be
\label{einstein}
S_{\rm g} =\frac{M^2_{\rm p}}{2}\int\dd^4x\,\sqrt{-g}\, \left( R - 6 H^2 \right) +  \frac{M^2_{\rm p} L^4}{2} \int\dd^4x\,\sqrt{-g}\,{W^{\mu\nu}}_{\rho\sigma}\, {W^{\rho\sigma}}_{\delta\tau}\, {W^{\delta\tau}}_{\mu\nu}~.
\ee
Here $M_{\rm p}$ is the Planck mass, $H$ is the Hubble scale, $W_{\mu\nu\rho\sigma}$ is the Weyl tensor, and $L$ is a parameter with dimension of length. Our dimensionless parameter $N$ is of the order $M_{\rm p}^2 /H^2$, which current data set to be  at least of the order of $10^{9}$. The contribution due to the Einstein term has been computed in \cite{maldacenaNG}.
The same three-point structures can be obtained from considerations of three-dimensional conformal field theory \cite{Osborn:1993cr,giombi}. There, the stress energy-momentum tensor
is constrained by conformal symmetries to be a linear combination of two structures (in the absence of parity violating terms): the stress tensor three-point
functions from free scalar and free fermion theories. 
One way to think about the scale $L$ in the context of inflation, is as an additional scale on top of the Planck and Hubble scales. Imagine, for instance, that $L$ is the string length in a weakly coupled string theory with $1 \ll L M_{\rm p}$. When the higher-spin string states are very massive with respect to $H$ we have $1 \gg HL \gg H/M_{\rm p}$. In another regime, where $1 \ll M_{\rm p} /H$ and $H L \lesssim 1$, the string states are light with respect to the Hubble scale during inflation (without losing perturbative control of the string). In this second scenario, we should consider a tower of light higher-spin particles in an approximate de Sitter background. In the absence of experimental data on the number $HL$, such a scenario seems phenomenologically viable. We will calculate $H L$ exactly for the Vasiliev theory to confirm that it falls under the second category. 
In other words, the higher-spin theory under consideration generates a different graviton three-point from that
of Einstein gravity \cite{ei}. The difference can be viewed as originating from a cubic Weyl term with a specific coefficient. 

That the early phase of the universe may be of the higher-spin type is an appealing scenario. However, we are still left with the important and open problem of describing the exit from inflation, as well as the generation of scalar fluctuations with the observed scalar tilt.  We do not resolve these issues, although we briefly offer some speculative remarks in the conclusions.

The paper is organised as follows. In section 2 we briefly review aspects of the $Q$-formalism. Section 3 contains the calculation of the two-point correlator of the spin-2 field. Section 4 is devoted to the calculation of the three-point function of the spin-2 field. Section 5 relates the three-point function to that stemming from an Einstein plus cubic Weyl term, and we provide the specific value of $HL$. In section 6 we discuss the shape of the tensor non-Gaussianities. Finally, in section 7 we offer some concluding remarks. The paper contains several appendices where technical details for the calculations can be found. 

\section{Review of the $Q$-formalism}

In this section we briefly summarise the proposal put forward in Ref. \cite{anninos}. The reader is invited to read this reference for more details.

\subsection{The free massless spin-$s$ in  de Sitter spacetime}
Let us consider a  de Sitter spacetime with conformal time $\eta$ and metric
\be
\dd s^2=\frac{1}{H^2\eta^2}\left(-\dd \eta^2+\dd\vx^2\right)~, \quad \vx \in \mathbb{R}^3~, \quad \eta \in (-\infty,0)~.
\ee
A free massless field of integer spin-$s$  is described a totally symmetric tensor $\phi_{\mu_1 \cdots \mu_s}(\eta,\vec x)$ obeying the free Fronsdal equation \cite{Fronsdal:1978rb, Fronsdal:1978vb}. Due to the gauge symmetry, the field is invariant under the following transformation
\begin{equation}
	\delta \phi_{\mu_1\cdots \mu_s} = \nabla_{(\mu_1}\Lambda_{\mu_2\cdots \mu_s)},
\end{equation}
with an arbitrary traceless gauge parameter,  
\be
\Lambda^\nu_{\,\,\nu \mu_3\cdots \mu_{s-1}}=0.
\ee
Moreover, for $s\ge 4$ the fields obey the double-traceless condition ${\phi^{\alpha \beta}}_{\alpha\beta\mu_1 \cdots \mu_{s-4}}=0$. The gauge invariance implies that each higher-spin field contains only two physical degrees of freedom.
In the gauge where one or more time indices are set to zero, the mode expansion of the $\phi_{i_1 \cdots i_s}(\eta,\vec x)$ reads\footnote{From now on we will use the following shorthand notations of Ref. \cite{anninos}
\be\nn
\int_k= \int \frac{\dd^3 k}{(2\pi)^3}, \qquad \delta_{\vk+\vk'}=(2\pi)^3\delta^{(3)}(\vk+\vk').
\ee
}
\begin{equation} \label{hsqf}
 \phi_{i_1 \cdots i_s}(\eta,\vx) = \sqrt{\gamma} \sum_{\lambda}  \int_k \left(a_{\vk}^{\lambda}  \, \psi^{ \lambda}_{\vec {k},i_1 \cdots i_s}(\eta) e^{i \vec {k} \cdot \vec x}  + {\rm h.c.}\right)~.
\end{equation}
Here, $\delta^{i_1 i_2} \psi^{ \lambda}_{\vec {k},i_1 i_2 \cdots i_s}= {k}^{i_1} \psi^{ \lambda}_{\vec {k},i_1 \cdots i_s} = 0$, and $\lambda$ indicates the helicity. The creation and annihilation operators satisfy the 
 canonical commutation relations 
\begin{equation}
 [ a_\vk^\lambda, a_{\vk'}^{\lambda'\dagger}] = \delta^{\lambda \lambda'} \, \delta_{\vk+\vk'}.
\end{equation} 
The parameter $\gamma$ is a normalisation factor which accounts for the fact that in the action the coefficient of the kinetic term of the field might not be 1/2, but rather $1/2\gamma$.
The vacuum two-point function in momentum space is 
\begin{align}
 \langle 0|\, {\phi}_{i_1 \cdots i_s}(\eta,\vk) \, {\phi}_{i'_1 \cdots i'_s}(\eta',\vk') \,  |0\rangle
  = \gamma(\eta \eta')^{\frac{3}{2}-s}   \, \frac{\pi}{4} \, 
 H^{(1)}_{s-\frac{1}{2}}\bigl(-k \eta\bigr) \, H^{(2)}_{s-\frac{1}{2}}\bigl(-k \eta'\bigr) \, \Pi_{i_1 \cdots i_s,i'_1 \cdots i'_s}(\vk) \, \delta_{\vk+\vk'}.
\end{align}
Here
$\Pi_{i_1 \cdots i_s,i'_1 \cdots i'_s}(\vk)$ indicates  the projector onto spin-$s$ transverse traceless polarisations (see Appendix\hspace{0.1cm}\ref{projector} for more details),
\begin{equation} \label{projector}
 \Pi_{i_1 \cdots i_s,i'_1 \cdots i_s'}(\vk) = \sum_\lambda \epsilon^\lambda_{i_1 \cdots i_s}(\vk) \,  \epsilon^{*\lambda}_{i'_1 \cdots i'_s}(\vk),
\end{equation}
and we have introduced the polarisation tensors  $\epsilon^\lambda_{i_1 \cdots i_s}(\vk)$ satisfying the following rule\footnote{We leave the basis choice free for the moment. In the  following we will specialise our results using  both the basis $X$ and $P$ adopted in Ref. \cite{mp} (where the polarisation tensors are real) and the chiral basis.}
\be\label{normepsilon}
\sum_{i_1 \cdots i_s} \epsilon^{*\lambda_1}_{i_1 \cdots i_s} \epsilon^{\lambda_2}_{ i_1 \cdots i_s} = \delta^{\lambda_1 \lambda_2}.
\ee
If we decompose the Hankel functions of the first kind as  $H^{(1)}_\nu(z) = J_\nu(z) + i Y_\nu(z)$, we can rewrite the expression for the modes as
\begin{align} \label{alphabetamodes}
 \phi_{i_1 \cdots i_s}(\eta,\vx) =  (-\eta)\int_k \,  \left( \alpha_{i_1 \cdots i_s}(\vk)  \, \bar{J}_{s-\frac{1}{2}}(-k\eta)   +  \widetilde\beta_{i_1 \cdots i_s}(\vk) \, \overline{Y}_{s-\frac{1}{2}}(-k\eta) \right) e^{i \vk \cdot \vx},
\end{align}
with
\begin{align}
 \bar{J}_\nu(z) \equiv \sqrt{\tfrac{\pi}{2}} \, z^{-\nu} J_{\nu}(z), \qquad 
 \overline{Y}_\nu(z) \equiv \sqrt{\tfrac{\pi}{2}} \, z^{-\nu} \,Y_{\nu}(z),
\end{align}
and
\begin{align} \label{commrel}
 \alpha_{i_1 \cdots i_s}(\vk) &=  \sqrt{\gamma}k^{s-\frac{1}{2}}\,\, \bigl(a_{\vk}^{ \lambda} \, \epsilon^\lambda_{i_1 \cdots i_s} + a_{\vk}^{  \dagger \lambda} \, \epsilon^{*\lambda}_{i_1 \cdots i_s} \bigr)/\sqrt{2},  \\
 \widetilde\beta_{i_1 \cdots i_s}(\vk) &= \sqrt{\gamma} k^{s-\frac{1}{2}} \, i \, \bigl(a_{\vk}^{\lambda} \, \epsilon^\lambda_{i_1 \cdots i_s} - a_{\vk} ^{\dagger \lambda} \, \epsilon^{*\lambda}_{i_1 \cdots i_s} \bigr)/\sqrt{2}.
\end{align}
The fields $\alpha_{i_1 \cdots i_s}(\vk) $ and $\widetilde\beta_{i_1 \cdots i_s}(\k)$ have conformal dimension $\widetilde \Delta={s+1}$. If we expand  at late times (i.e. $\eta \to 0)$ we arrive at\footnote{Note that the spin-2  graviton is identified here as the perturbation of the metric as $\dd s^2=\left(-\dd \eta^2+\dd\vx^2\right)/H^2\eta^2+h_{ij}(\eta,\vx) \dd x^i \dd x^j/H^2$. In the following we will use the more standard cosmology notation and identify the graviton with the perturbation of the spatial part of the metric as $\gamma_{ij}(\eta,\vx) \dd x^i \dd x^j/H^2\eta^2$.}
\begin{equation} \label{futurephi}
 \phi_{i_1 \cdots i_s}(\eta,\vx) \approx  \int_k  \,  \left( c_1 \, \alpha_{i_1 \cdots i_s}(\vk) \,  \eta   +  c_2 \,  k^{1-2s} \, \widetilde\beta_{i_1 \cdots i_s}(\vk) \, \eta^{2-2s}  \right) e^{i \vk \cdot \vx},
\end{equation}
with $c_1=-\sqrt{\pi/2}(1/2^\nu\Gamma(\nu+1))$ and $c_2=-\sqrt{\pi/2}2^\nu\Gamma(\nu)/\pi$, $\nu=s-1/2$.
One can define  a boundary field $\beta(\vx)$ related to $\widetilde\beta(\vx)$ by the following transformation in momentum space
\begin{align} \label{hsshadowrel}
 \widetilde\beta_{i_1 \cdots i_s}(\vk) = k^{2s-1} \, \beta_{i_1 \cdots i_s}(\vk) = \int_{k'} G_{i_1 \cdots i_s,i_1' \cdots i_s'}(\vk,\vk') \, \beta_{i_1' \cdots i_s'}(\vk'),
\end{align} 
where
\begin{align} \label{Gi1isip1sips}
 G_{i_1 \cdots i_s,i_1' \cdots i_s'}(\vk,\vk')= k^{2s-1} \, \Pi_{i_1 \cdots i_s,i_1' \cdots i_s'}(\vk) \, \delta_{\vk+\vk'}.
\end{align}
Since   $G_{i_1 \cdots i_s,i_1' \cdots i_s'}$ represents  the two-point function of a spin-$s$ conserved current, this can be interpreted as the CFT shadow transform, which we discuss in Appendix\hspace{0.1cm}\ref{shadow}. As a consequence, the field  $\beta(\vx)$ transforms as a spin-$s$ primary field with conformal dimension $\Delta=2-s$, while $\widetilde \beta(\vx)$ transforms as a spin-$s$ primary field of conformal dimension $\widetilde{\Delta}=s+1$, with $\Delta+\widetilde{\Delta}=3$.

\subsection{Microscopic operator content}

In Ref. \cite{anninos} it was argued that the microscopic operator content of the theory is given by $2N$ Hermitian operators $\hat{Q}^\alpha(\vx)$, and other additional $2N$ Hermitian operators $\hat{\Pi}^\alpha(\vx)$, with $\alpha=1,\cdots,2N$. In the semi-classical regime the dimensionless parameter $N$, which goes as $M_{\rm p}^2/H^2$, is very large. The microscopic operators satisfy the algebra
\begin{equation}
[\hat{Q}^\alpha(\vx),\hat{\Pi}^\beta(\vy) ] = i \delta(\vec{x}-\vec{y}) \delta^{\alpha\beta}~,
\end{equation}
with all other commutators vanishing. From now on, we will work in the $\hat{Q}$-eigenbasis, in which we can identify the late-time Bunch-Davies quantum state with the Gaussian wavefunctional:
\begin{equation}
\Psi(Q) = \exp \left[\frac{1}{2} \int_x Q^\alpha(\vec{x}) \partial_{\vec{x}}^2 \, Q^\alpha(\vec{x})\right]~.
\end{equation}
The integration measure for the $Q$ is the standard flat measure. The Gaussian structure of $\Psi$ allows us to calculate all expectation values using relatively simple Wick contractions. 

\begin{figure}[t]
	\includegraphics[width=0.8 \linewidth]{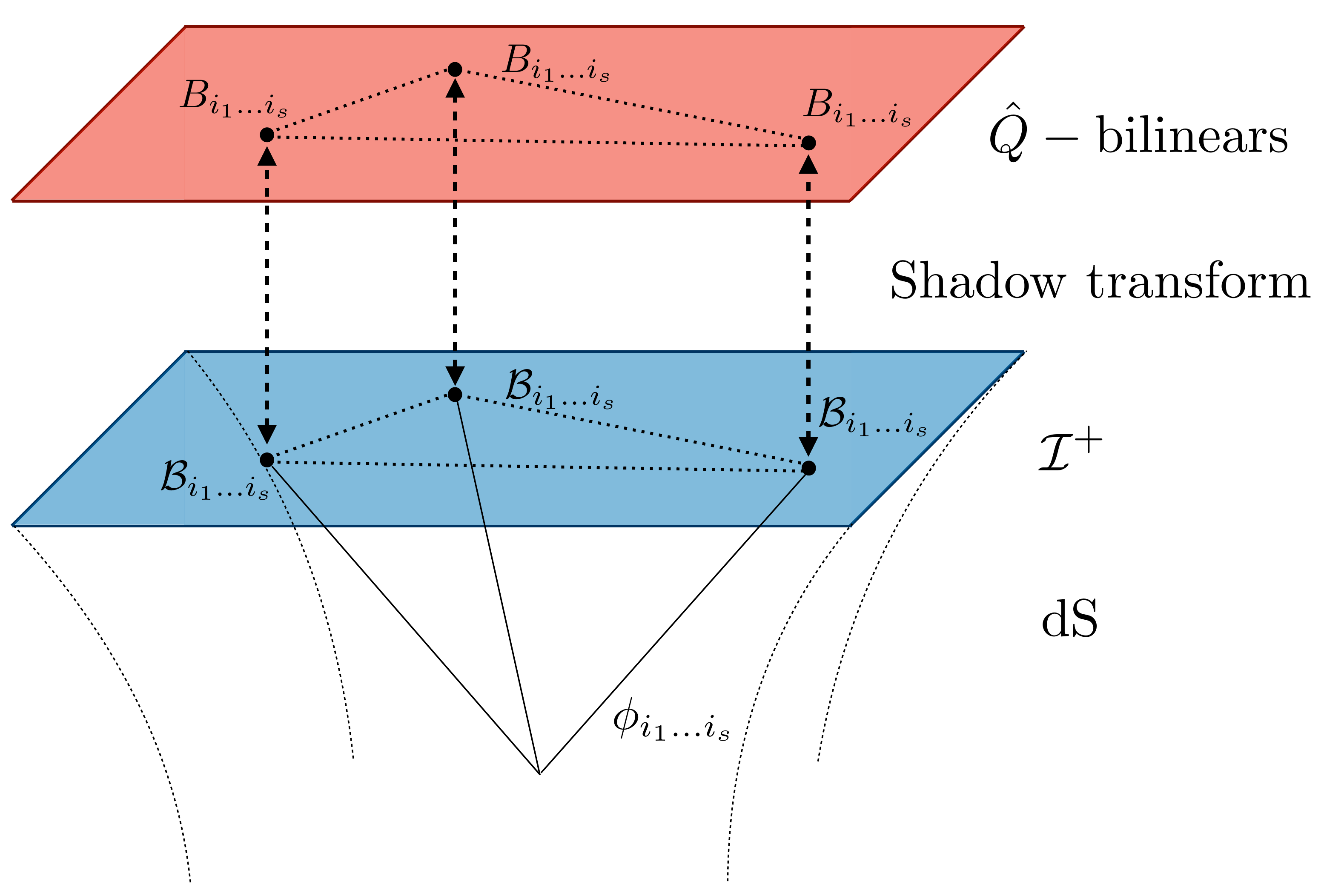}
	\centering
	\caption{Pictorial representation of the identification between bulk effective fields and the microscopic bilinear operators. $\mathcal{I}^+$ denotes the future boundary of de Sitter space.}
	\label{fig:dS/CFT}
\end{figure}

The higher-spin boundary fields  $\tilde{\beta}_{i_1 \cdots i_s}(\vx)$ are identified with the bilinear operators 
 \be
 B_{s,i_1 \cdots i_s}(\vx) = \frac{1}{N} \int \dd^3 y: Q^\alpha(\vx) {\cal D}_{i_1 \cdots i_s}(\vx,\vy) Q^\alpha(\vy):~, 
 \ee
 where the differential operator $ {\cal D}_{i_1 \cdots i_s}$ are those appearing in the construction of conserved currents in the free O(2$N$) model. The positive integer $N$ is proportional to the square of the ratio between the Planck mass and the Hubble rate. The symbol $:\ :$ represents normal ordering.
 It will be convenient to introduce the shadow transform $\CB_{s, i_1' \cdots i_s'}$ of the bilinear $B_{s, i_1 \cdots i_s}$
 \be
 B_{s, i_1 \cdots i_s}(\vk)  = \int_{k'} G_{i_1 \cdots i_s,i_1' \cdots i_s'}(\vk,\vk') \CB_{s, i_1' \cdots i_s'}(\vk'),
 \ee
It then follows that one can identify the boundary higher-spin profiles $\beta_{ i_1 \cdots i_s}$ with the bilinears $\CB_{s, i_1 \cdots i_s}$. In Fig.~\ref{fig:dS/CFT} we show a schematic representation of the various relations.

 In order to  efficiently  generate the  higher-spin currents constructed out from the 2$N$ bosonic real fields  $Q^\alpha(\vx)$, one makes use of  the equivalence between traceless symmetric tensors and functions of a complex null vector $z$ to build up the scalar
\begin{equation}
B_s(\vx|z) \equiv B_{s, i_1 \ldots i_s}(\vx) z^{i_1} \ldots z^{i_s} \propto \sum_{k=0}^s a^{(s)}_k(z\!\cdot\!\partial)^k Q^\alpha(\vx)(z \cdot \partial)^{s-k}Q^\alpha(\vx),
\end{equation}
which should be restricted to the submanifold $z^2 = 0$.
The coefficients $a_k$ of the expansion are defined by the rule
\begin{equation}
\sum_{k=0}^sa_k^{(s)}x^ky^{s-k}=f_s(x,y),
\end{equation}
where one  chooses
\begin{equation}
f_s(x,y)= \frac{2^{(5-s)/2}}{N} (x+y)^sT_s\left(\frac{x-y}{x+y}\right)
\end{equation}
in order to normalise properly the two-point functions. Here $T_s(u)$ represents the Chebyshev polynomial of order $s$.

As an example, the spin-2 field $B_2(\vx|z)$ can be written as \cite{Sleight:2016dba,anninos}
\begin{equation}
B_2(\vx|z)=\frac{2\sqrt{2}}{N}:\!\left[(z\!\cdot\!\partial)^2Q^{\alpha}(\vx)Q^{\alpha}(\vx)+Q^{\alpha}(\vx)(z\!\cdot\!\partial)^2Q^{\alpha}(\vx)-6(z\cdot\partial)Q^{\alpha}(\vx)(z\!\cdot\!\partial)Q^{\alpha}(\vx)\right]\!:~.
\end{equation}
In general, the correlator of two higher-spin fields $B_s$ with these normalisations becomes
\begin{equation}
\label{2pps}
    \langle B_s(\vx_1|z_1)B_{s'}(\vx_2|z_2)\rangle=\frac{(2s)!}{\pi^2N}\frac{(z_1\!\cdot\!H(x_{12})\cdot\! z_2)^s}{(x_{12}^2)^{1+s}} \delta^{ss'}~,
\end{equation}
valid for $ s \ge 1$. Here 
 $H_{ij}(x) = \delta_{ij} - 2 x_i x_j / x^2$  and $x_{ij} = \vx_i - \vx_j$. For scalar and spin-2 operators,  the two-point functions are of the form
\begin{equation}
    \langle B_0(\vx_1)B_0(\vx_2)\rangle=\frac{1}{N}\frac{1}{2\pi^2 x_{12}^2},\quad   \langle B_2(\vx_1|z_1)B_{2}(\vx_2|z_2)\rangle=\frac{1}{N}\frac{4!(z_1\!\cdot\! H(x_{12})\!\cdot\! z_2)^2}{\pi^2x_{12}^6}~.
\end{equation}
The equation above also fixes the normalisation coefficient $\gamma$ to be $2/N$. In momentum space, the polynomials are Fourier transformed and, for example, spin-zero and spin-2 fields take the following expression
\begin{eqnarray}
\label{Bexpan}
B_0(p) &=& \frac{2\sqrt{2}}{N}\int_{q} \,:\!Q^\alpha_q Q^{\alpha}_{p-q}\!: , \\
B_2(p|z) &=& -\frac{2\sqrt{2}}{N}\int_{q} :\!Q^\alpha_{q}Q^{\alpha}_{p-q}\!: [(z\!\cdot\! q)^2+(z\!\cdot\!({p-q}))^2+6 (z\!\cdot\!  q) z\!\cdot\!(q- p)]. \label{Bexpan2}
\end{eqnarray}
One can go back and find the original tensors correlator by factorising and dropping the $z$ vectors. For example, the graviton three-point function can be written as
\begin{equation}\label{polexpBBB}
\langle B_{2}(p_1|z_1)B_{2}(p_2|z_2)B_{2}(p_3|z_3)\rangle = \langle B_{2,ab}(p_1)B_{2,ij}(p_2)B_{2,mn}(p_3)\rangle z_1^az_1^b z_2^i z_2^j z_3^m z_3^n.
\end{equation}
Once these boundary correlators have been calculated, one has to apply the shadow transform explained in Appendix\hspace{0.1cm}\ref{shadow}.

\subsection{Relation to dS/CFT}

In the context of higher-spin gravity, the $Q$-formalism can be viewed as a completion of the proposal \cite{maldacenaNG} that the late-time de Sitter invariant Euclidean wavefunctional $\psi_0$ \cite{Chernikov:1968zm,Boerner:1969ff,Candelas:1975du,Dowker:1975tf,Schomblond:1976xc,BD,Hartle:1983ai,Anninos:2014lwa} is calculated by a CFT partition function. At the semi-classical level, $\psi_0$ is a functional of the late-time profiles $\beta_{i_1\ldots i_s}(\vec{x})$ of all the higher-spin fields. From the dual CFT perspective, the $\beta_{i_1\ldots i_s}(\vec{x})$ are understood as sources for each conserved current operator \cite{stromanninos}. In order to calculate actual expectation values, we must integrate $|\psi_0|^2$ over all these profiles using an integration measure invariant under all the higher-spin gauge symmetries. In the semi-classical limit, and at low enough order in perturbation theory, this measure can be taken to be flat. More generally, however, the assumption of a flat measure need not hold. The original dS/CFT proposal does not provide an integration measure. In calculating physical expectation values of bulk fields, the $Q$-formalism implicitly provides an exact integration measure, thus completing the original dS/CFT proposal. Wherever a direct comparison is available, such as the semi-classical or perturbative limit, the two proposals are in complete agreement. Moreover, the $Q$-formalism extends the dS/CFT story by providing a complete Hilbert space and microscopic operator content for the bulk theory, as opposed to the description of a single quantum state.

\section{Two-point correlator of the spin-2 field} 

We are now ready to compute  the graviton correlators. In this section we consider the two-point function in momentum space. We present it since it is simpler than the calculation of the three-point correlator and 
we can highlight some of the technical details involved. These will be useful when dealing with the more involved calculation of the three-point correlator.
\subsection{Real space}
One straightforward way to get the   two-point correlator of the spin-2 field is to 
 start from its  expression  in position space,  given by  Eq. \eqref{2pps}
\begin{equation}
	 \langle B_2(\vx_1|z_1)B_{2}(\vx_2|z_2)\rangle=\frac{1}{N}\frac{4!(z_1\!\cdot\! H(x_{12})\!\cdot\! z_2)^2}{\pi^2x_{12}^6}.
\end{equation}
Performing the Fourier transformation  to express the expectation value in  momentum space and picking up the null vectors $z_1$ and $z_2$, we obtain  
\be
\langle B_{2,ij}(p_1)B_{2,lm}(p_2)\rangle= \frac{1}{N}p_1^{3}\Pi_{ij,lm}( p_1)\delta_{\vec{p}_1+\vec{p}_2}.
\ee
We  can now apply the shadow transform to obtain the  local boundary fields as shown in Eq. \eqref{BCB} to get 
\begin{equation}
\label{real}
\langle \CB_{2, i j}(p_1)\CB_{2,  lm}(p_2)\rangle =
 \frac{1}{p_1^3p_2^3} {\Pi^{i'j'}}_{,ij}({\vec p_1}){\Pi^{ l' m'}}_{,lm}({\vec p_2}) 
 \langle B_{2, i' j'}(p_1)B_{2, l' m' }(p_2)\rangle =
 \frac{1}{N}\frac{1}{p_1^3}\Pi_{ij,lm}( \vec p_1)\delta_{\vec{p}_1+\vec{p}_2}. 
\end{equation}
This equation provides the expression for the two-point function of the spin-2 field  in momentum space.
We  now turn to the computation of the same expectation value through the formalism outlined in the previous section.

\subsection{Momentum space}
The two-point function  of the spin-2 field  can be obtained starting from the expression of the  spin-2 field  in Eq. \eqref{Bexpan}
\begin{align}
\label{grav2}
\langle B_2(p_1|z_1)B_2(p_2|z_2)\rangle &=\frac{8}{N^2}\int_{k_1}\int_{k_2}\bigl\langle \, :\!Q_{k_1}Q_{p_1-k_1}\!: \, :\!Q_{k_2}Q_{p_2-k_2}\!: \bigr\rangle \nonumber \\
&\times [(z_1\cdot k_1)^2+(z_1\cdot(p_1-k_1))^2+6(z_1\cdot k_1)z_1\cdot(k_1-p_1)] \nonumber \\
&\times [(z_2\cdot k_2)^2+(z_2\cdot(p_2-k_2))^2+6(z_2\cdot k_2)z_2\cdot(k_2-p_2)].
\end{align}
We explicitly compute the normal ordered expectation value as
\begin{eqnarray}
\label{wick2}
		\bigl\langle \, :\!Q_{k_1}Q_{p_1-k_1}\!: \, :\!Q_{k_2}Q_{p_2-k_2}\!: \bigr\rangle
		&=&
		N  \color{black} k_1^{-2} k_2^{-2} \delta_{\vk_1 +\vp_2 -\vk_2}\delta_{\vk_2 +\vp_1-\vk_1}.
\end{eqnarray}
Thus,
\begin{align}
\langle B_2(p_1|z_1)B_2(p_2|z_2)\rangle &=\frac{8}{N}\int_{k_1}\int_{k_2}\, k^{-2}_1 k^{-2}_2 \delta_{\vec{p}_1-\vec{k}_1+\vec{k}_2}\delta_{\vec{p}_2-\vec{k}_2+\vec{k}_1} \nonumber \\
&\times [(z_1\cdot k_1)^2+(z_1\cdot(p_1-k_1))^2+6(z_1\cdot k_1)z_1\cdot(k_1-p_1)] \nonumber \\
&\times [(z_2\cdot k_2)^2+(z_2\cdot(p_2-k_2))^2+6(z_2\cdot k_2)z_2\cdot(k_2-p_2)] .
\end{align}
The result depends  on a minimal set of integrals as follows
\begin{eqnarray}\label{22.a}
\langle B_2(p_1|z_1)B_2(p_2|z_2)\rangle &=&\frac{8}{N}\delta_{\vec{p}_1 + \vec{p}_2} (A_{z_1, z_2}+B_{z_1, z_2}+C_{z_1, z_2}+C_{z_2, z_1} \nonumber \\
&+&6D_{z_1, z_2}+6D_{z_2, z_1}+6E_{z_1, z_2}+6E_{z_2, z_1}+36 F_{z_1, z_2}),
\end{eqnarray}
where we have defined 

\begin{align}
 A_{z_1, z_2}&=z_{1  i} z_{1  j} z_{2 l} z_{2 m}
\int_{k_1} \frac{1}{|\vec{k}_1|^2} \frac{1}{|\vec{k}_1 - \vec{p}_1|^2}
  \lp k_1^i k_1^j k_1^l k_1^m \rp
\equiv {z_{1  i} z_{1  j} z_{2 l} z_{2 m} I_A^{i j l m}},
\nonumber \\
 B_{z_1, z_2}&=z_{1  i} z_{1  j} z_{2 l} z_{2 m}
\int_{k_1} \frac{1}{|\vec{k}_1|^2} \frac{1}{|\vec{k}_1 - \vec{p}_1|^2}
[ \lp k_1-p_1 \rp^i \lp k_1-p_1 \rp^j \lp k_1-p_1 \rp^l \lp k_1-p_1 \rp^m]
\nn
\\
&\equiv {z_{1  i} z_{1  j} z_{2 l} z_{2 m} I_B^{i j l m}},
\nonumber \\
C_{z_1, z_2}&=z_{1  i} z_{1  j} z_{2 l} z_{2 m}
\int_{k_1} \frac{1}{|\vec{k}_1|^2} \frac{1}{|\vec{k}_1 - \vec{p}_1|^2}
[ k_1^i k_1^j  \lp k_1-p_1 \rp^l \lp k_1-p_1 \rp^m]
\equiv {z_{1  i} z_{1  j} z_{2 l} z_{2 m} I_C^{i j l m}},
\nonumber \\
 D_{z_1, z_2}&=z_{1  i} z_{1  j} z_{2 l} z_{2 m}
\int_{k_1} \frac{1}{|\vec{k}_1|^2} \frac{1}{|\vec{k}_1 - \vec{p}_1|^2}
 [ k_1^i\lp k_1-p_1 \rp^j k_1^l k_1^m ] 
 \equiv {z_{1  i} z_{1  j} z_{2 l} z_{2 m} I_D^{i j l m}},
\nonumber \\
 E_{z_1, z_2}&=z_{1  i} z_{1  j} z_{2 l} z_{2 m}
\int_{k_1} \frac{1}{|\vec{k}_1|^2} \frac{1}{|\vec{k}_1 - \vec{p}_1|^2}
[  k_1^i \lp k_1-p_1 \rp^j \lp k_1-p_1 \rp^l \lp k_1-p_1 \rp^m ]
\equiv {z_{1  i} z_{1  j} z_{2 l} z_{2 m} I_E^{i j l m}},
\nonumber \\
 F_{z_1, z_2}&=z_{1  i} z_{1  j} z_{2 l} z_{2 m}
\int_{k_1} \frac{1}{|\vec{k}_1|^2} \frac{1}{|\vec{k}_1 - \vec{p}_1|^2} 
[ k_1^i \lp k_1-p_1 \rp^j  k_1^l \lp k_1-p_1 \rp^m]
\equiv {z_{1  i} z_{1  j} z_{2 l} z_{2 m} I_F^{i j l m}}.
\end{align}
We notice that we only have five types of integrals, since $I_F^{i j l m}=I_C^{i  l j m}$. Furthermore, one can prove that $I_B^{i j l m}=I_A^{i  j  l m}$  by shifting  $\vec k_1 \to \vec k_1 = \vec k_1 - \vec p_1 $ and then changing its sign.
Along the same lines, one can prove that $I_E^{i j l m}=I_D^{j  i  l m}$ leaving  finally only three integrals to evaluate, namely
\begin{align}
& I_1^{i j l m}=
\int_{k_1} \frac{1}{|\vec{k}_1|^2} \frac{1}{|\vec{k}_1 - \vec{p}_1|^2}
\lp  k_1^i k_1^j k_1^l k_1^m\rp, 
\nonumber \\
&I_2^{i j l m}=
\int_{k_1} \frac{1}{|\vec{k}_1|^2} \frac{1}{|\vec{k}_1 - \vec{p}_1|^2}
[ k_1^i k_1^j  \lp k_1-p_1 \rp^l \lp k_1-p_1 \rp^m],
\nonumber \\
&I_3^{i j l m}=
\int_{k_1} \frac{1}{|\vec{k}_1|^2} \frac{1}{|\vec{k}_1 - \vec{p}_1|^2}
[ k_1^i\lp k_1-p_1 \rp^j k_1^l k_1^m ].
\end{align}
The expression \eqref{22.a} becomes
\begin{align}\label{22.b}
&\langle B_2(p_1|z_1)B_2(p_2|z_2)\rangle =
\frac{8}{N}\delta_{\vec{p}_1 + \vec{p}_2} 
z_{1  i} z_{1  j} z_{2 l} z_{2 m}
  \lp
2 I_1^{i j l m}+ I_2^{i j l m} +I_2^{ l m i j } + 36 I_2 ^{i  l j m} + 12  I_3^{i j l m} +12 I_3^{ l m i j } 
  \rp. 
\end{align}
Since we are left with a single integrated momentum, we redefine  $\vec k_1 \equiv \vec k$ and we expand the numerators in terms of their powers of $\vec p_1$ to get
\begin{equation}
 I_1^{i j l m} =
\mathcal{I}_0^{i j l m}, \quad\quad
I_2^{i j l m} =
\mathcal{I}_0^{i j l m} - \mathcal{I}_1^{i j l m}- \mathcal{I}_1^{i j m  l}+\mathcal{I}_2^{i j l m},
 \quad\quad
I_3^{i j l m} =
\mathcal{I}_0^{i j l m} - \mathcal{I}_1^{i  m  l j},
\end{equation}
where we have defined:
\begin{align}
& \mathcal{I}_0^{i j l m}=
\int_{k} \frac{1}{|\vec{k}|^2} \frac{1}{|\vec{k} - \vec{p}_1|^2}
 \lp  k^i k^j k^l k^m \rp,
\nonumber \\
& \mathcal{I}_1^{i j l m}=
\int_{k} \frac{1}{|\vec{k}|^2} \frac{1}{|\vec{k} - \vec{p}_1|^2}
 \lp  k^i k^j k^l p_1^m \rp ,
\nonumber \\
& \mathcal{I}_2^{i j l m}=
\int_{k} \frac{1}{|\vec{k}|^2} \frac{1}{|\vec{k} - \vec{p}_1|^2}
\lp  k^i k^j p_1^l p_1^m  \rp .
\end{align}
Finally, we arrive at the following minimal expression in terms of the integrals $\mathcal{I}_i$
\begin{align}\label{22.c}
\langle B_2(p_1|z_1)B_2(p_2|z_2)\rangle 
  &= \frac{8}{N}\delta_{\vec{p}_1 + \vec{p}_2} \nn
z_{1  i} z_{1  j} z_{2 l} z_{2 m}
  \left[ 
  \lp 
  15 \I_0^{i j l m}+ 13 \I_0^{ l m i j }+ 36 \I_0^{i  l j m   }
  \rp \right .
 \\ \nn
 &\left . - \lp
2  \I_1^{i j l m} +2 \I_1^{l m i j }
+36  \I_1^{i l m  j } +12  \I_1^{i m   l j } +12  \I_1^{l j i m    } 
 \rp 
 \right .
 \\
 &\left . +\lp 
 \I_2^{ij l   m} +\I_2^{ l   m ij } +36 \I_2^{ i l  j m  }
  \rp
  \right].
 \end{align}
\subsubsection{Evaluating the integrals and their regularisation}\label{symarg}
The previous formula (\ref{22.c}) looks rather cumbersome. At this point, it is convenient to  consider the final step one should perform to compute the  two-point function  of the spin-2 field, that is  the shadow transform of the correlator of $B_{ij}$ to a correlator of $\CB_{ij}$. 
This leads to a simplification of the expression (\ref{22.c}). The argument goes as follows. We know that the correlator of the boundary fields we are interested in can be found by shadow transforming \eqref{22.c} as
\begin{equation}
\label{ShadowBB}
\langle \CB_{2, i j}(p_1)\CB_{2, k l}(p_2)\rangle = \frac{1}{ p_1^3p_2^3}
 {\Pi^{i'j'}}_{,ij}({\vec p_1})
 {\Pi^{k' l'}}_{,kl}({\vec p_2}) \langle B_{2, i' j'}(p_1)B_{2,k' l'}(p_2)\rangle,
\end{equation}
where we have used the inverse of $B_{2, ij}(p)=p^3 \, {\Pi^{i'j'}}_{,ij}({\vec p}) \,\CB_{2, i'j'}(p)$. Once the fictitious contractions with vectors $z_i$ are eliminated, what we are left with is a tensor contraction of propagators ${\Pi^{i'j'}}_{,ij}({\vec p_1}){\Pi^{k' l'}}_{,kl}({\vec p_2})$ with a tensor $T_{i'j'k'l'}$ build out of $\vec p_1$, $\vec p_2$ and $\delta_{ij}$. 
The most general tensor which can be contracted with two projectors, giving a projector as a result, is a projector itself. Namely,
\begin{equation}
	{\Pi^{i'j'}}_{,ij}({\vec p}){\Pi^{k' l'}}_{,kl}({\vec p}) \Pi_{i' j', k' l'} (\vec p)= \Pi_{ij,kl}(\vec p),
\end{equation}
where we used the fact that, in a two point function, the moduli satisfy the relations $p_1=p_2=p$. This particular contraction is the only structure which allows for momenta in $T_{i'j'k'l'}$.
No other contractions with the momenta are possible since the projector is transverse\footnote{
For example, for the  spin-one state  (the generalisation to HS is trivial due to the properties of the projector tensors) one can see that
$$
\Pi_{ij}\Pi_{kl}\Pi^{jk}=
\lp \delta_{ij} - \hat p_i \hat p_j\rp
 \lp \delta_{kl} - \hat p_k \hat p_l\rp
 \lp \delta^{jk} - \hat p^j \hat p^k\rp
 =
 \lp \delta_{ij} - \hat p_i \hat p_j\rp
  \lp \delta^{j}_l - \hat p^j\hat p_l\rp
  =
  \lp \delta_{il} - \hat p_i \hat p_l\rp
  =\Pi_{il}.
$$
Notice that the same result is achieved by considering only the contraction with the delta such that
$$
\Pi_{ij}\Pi_{kl}\delta^{jk}=
\lp \delta_{ij} - \hat p_i \hat p_j\rp
 \lp \delta_{kl} - \hat p_k \hat p_l\rp
 \delta^{jk} 
 =
 \lp \delta_{ij} - \hat p_i \hat p_j\rp
  \lp \delta^{j}_l - \hat p^j\hat p_l\rp
  =
  \lp \delta_{il} - \hat p_i \hat p_l\rp
  =\Pi_{il}.
$$}.
We can extract therefore the  overall coefficient by computing only the terms proportional to the Kronecker deltas,
namely
\begin{equation}
{\Pi^{i'j'}}_{,ij}({\vec p_1}){\Pi^{k' l'}}_{,kl}({\vec p_2}) \lp \delta_{i' k'} \delta_{j' l'}  \rp  
\qquad \text{or}  \qquad  
{\Pi^{i'j'}}_{,ij}({\vec p_1})
{\Pi^{k' l'}}_{,kl}({\vec p_2}) \lp \delta_{i' l'} \delta_{j' k'}  \rp 
\end{equation}
and  isolate the term proportional to $(z_1 \cdot z_2 )^2$ coming from integrals $\I_i$ which are fully proportional to Dirac deltas. It is straightforward to see that only $\I_0$ remains 
\begin{equation}\label{B2B2}
\langle B_2(p_1|z_1)B_2(p_2|z_2)\rangle \supset \frac{8}{N}\delta_{\vec{p}_1 + \vec{p}_2} 
z_{1  i} z_{1  j} z_{2 l} z_{2 m}
  \lp 
  15 \I_0^{i j l m}+ 13 \I_0^{ l m i j }+ 36 \I_0^{i  l j m   }
  \rp,
\end{equation}
where the inclusion symbol  means that, as we  just mentioned, we are highlighting only the terms proportional to Kronecker deltas.

The integral $\mathcal{I}_0$ is divergent. One way to proceed is to compute the two-point function in real space and then Fourier transform it \cite{Bzowski:2013sza, Bzowski:2015pba,anninos,Bzowski:2018fql}, as we did in the previous section.  Otherwise one can regularise the integral and find the solution by using analytical continuation. More details are provided in 
 Appendix\hspace{0.1cm}\ref{doublek} and \ref{app:reg}.
We define the integral 
\be
\mathcal{I}_0^{i j l m}(d,\delta_1,\delta_2)=&
\int \frac{d^d \vec{k}}{(2\pi)^d} \frac{1}{|\vec{k}|^{2 \delta_1}} \frac{1}{|\vec{k} - \vec{p}_1|^{2 \delta_2}}
\lp  k^i k^j k^l k^m \rp \\
\supset & \frac{2^{{-d} + \delta_t -3} \,  {\rm S}^{i j l m} }{ \lp  4 \pi \rp  ^{\frac{d}{2}}  \Gamma(\delta_1) \Gamma(\delta_2) \Gamma \lp \frac{3d}{2} +6 -2 \delta _t \rp } 
I_{d+3-\delta_t\{d-2 \delta _t +\delta _1 +4, d-2 \delta _t +\delta _2 +4\}} (p_1,p_2),
\ee
where $\delta_t=\delta_1+\delta_2$ and 
\begin{equation}
I_{d+3-\delta_t\{d-2 \delta _t +\delta _1 +4, d-2 \delta _t +\delta _2 +4\}} (p_1,p_2)=	\int_0 ^\infty {\rm d} x\,
x ^{d+3 -\delta_t}
 \prod_{i=1}^{2}
 p_i^{d -2\delta _t +\delta _i +4}K_{d-2 \delta _t +\delta _i +4}(x p_i).
\end{equation}
We recall that the moduli $p_1=p_2$ due to the momentum conservation, so we define $p=p_1=p_2$. 
One can define the following coefficients to match the definition used in Eq.~\eqref{dkdef}
\be
\alpha &= d+3 -\delta_t ,
\\
\beta_1 &= d-2 \delta _t +\delta _1 +4,
\\
\beta_2 &=d-2 \delta _t +\delta _2 +4.
\ee
The regularisation method prescribes a shift of  the weights such that (we leave $u$ and $v$ undetermined following the notations used in Ref.~\cite{Bzowski:2013sza} and explained in Appendix\hspace{0.1cm}\ref{app:reg})
the solution of the double-K integral is (see Eq.~\eqref{e:I2K})
\be
&I_{\alpha+u\epsilon \{\beta_1 +v\epsilon, \beta_2 +v\epsilon \}} (p,p) =
  \frac{2^{\alpha - 2 +u\epsilon}}
{\Gamma(\alpha+1 +u\epsilon) p^{\alpha +1-\beta_1-\beta_2+u\epsilon-2 v\epsilon}}
\Gamma \left( \frac{\alpha + \beta_1 + \beta_2+1 +u\epsilon +2 v \epsilon}{2} \right) 
\\
& \Gamma \left( \frac{\alpha + \beta_1 - \beta_2+1+u\epsilon}{2} \right) 
\Gamma \left( \frac{\alpha- \beta_1 + \beta_2+1 +u\epsilon}{2} \right) 
\Gamma \left( \frac{\alpha - \beta_1 - \beta_2+1 +u\epsilon-2 v \epsilon}{2} \right)
\\
&=
\frac{2^{2+ u \epsilon}}{\Gamma(5+ u \epsilon)} p^{3- u \epsilon+2 v \epsilon} 
\Gamma \lp\frac{13}{2}+ \frac{ (u+2v)   \epsilon}{2}\rp 
\Gamma \lp\frac{5}{2}+ \frac{ u \epsilon}{2}\rp ^2
\Gamma \lp-\frac{3}{2}+ \frac{(u-2v)  \epsilon}{2}\rp, 
\ee
where we have chosen $d=3$, $\delta_1 =1$, $\delta_2 =1$ and $\delta _t =2$. Sending  $\epsilon \to 0$  we get 
\be
I_{4,\{ 4,4\}}(p,p)=\frac{10395 \pi ^2}{512}p^{3}
\ee
so that 
\be
\label{a}
\mathcal{I}_0^{i j l m} (3,1,1)\supset 
\frac{2^{-4} \,  {\rm S}^{i j l m} }{ \lp  4 \pi \rp  ^{\frac{3}{2}} \frac{10395 \sqrt{\pi }}{64} } 
I_{4,\{ 4,4\}}(p,p)
= 
2^{-10}p^{3} {\rm S}^{i j l m}.
\ee

\subsubsection{Final result}
By inserting  the expression (\ref{a}) into Eq. \eqref{B2B2} we finally arrive at
\begin{eqnarray}
\langle B_2(p_1|z_1)B_2(p_2|z_2)\rangle &\supset&  \frac{8}{N}\delta_{\vec{p}_1 + \vec{p}_2} 
z_{1  i} z_{1  j} z_{2 l} z_{2 m}
  2^{-10} p^3
  \lp 
  15 \, {\rm S}^{i j l m}+ 13\, {\rm S}^{ l m i j }+ 36\, {\rm S}^{i  l j m   }
  \rp\nonumber\\
&=&\frac{8}{N}\delta_{\vec{p}_1 + \vec{p}_2} 
z_{1  i} z_{1  j} z_{2 l} z_{2 m}
  2^{-4} p^3
   \lp \delta ^{i j}\delta ^{ l m} +\delta ^{il}\delta ^{  j m}+\delta ^{im}\delta ^{ j l } \rp.
\end{eqnarray}
 Using the fact that only the terms proportional to  $(z_1 \cdot z_2)^2$ survive since the other possible contractions are zero (since $z_i$ are null, \textit{i.e.} $z_i \cdot z_i=0$, the term proportional to $\delta ^{i j}\delta ^{ l m}$ does not survive),  we are left with 
  \be
\langle B_2(p_1|z_1)B_2(p_2|z_2)\rangle \supset \frac{p^3}{N}\delta_{\vec{p}_1 + \vec{p}_2} 
z_{1  i} z_{1  j} z_{2 l} z_{2 m}
\frac{1}{2}
\delta ^{il}\delta ^{  j m} .
\ee
Using now the polynomial prescription we find
 \be
\langle B_{2, i j }(p_1)B_{2, l  m }(p_2)\rangle  \supset \frac{p^3}{N}\delta_{\vec{p}_1 + \vec{p}_2} 
\frac{1}{2}
\delta _{il}\delta _{  j m}  ,
\ee
which  corresponds to the first term of the reconstructed projector tensor. In fact, if one considers all the other possible terms (we do not report here the lengthy, but straigthfoward calculation),  one obtains
 \be
\langle B_{2, i j }(p_1)B_{2, l  m }(p_2)\rangle = \frac{p^3}{N} \Pi _{il , j m} (\vec p_1) \delta_{\vec{p}_1 + \vec{p}_2} 
.
\ee
Shadow transforming the previous formula, we find
\begin{eqnarray}
\langle \CB_{2, i j}(p_1)\CB_{2, l m}(p_2)\rangle &=& 
\frac{1}{p_1^3p_2^3}
 {\Pi^{i'j'}}_{,ij}({\vec p_1})
{\Pi^{ l' m'}}_{,l m }({\vec p_2}) 
\langle B_{2, i' j'}(p_1)B_{2,  l' m'}(p_2)\rangle \nonumber\\
&=&
\frac{1}{N}
\frac{1}{p_1^3}
 {\Pi^{i'j'}}_{,ij}({\vec p_1})
{\Pi^{ l' m'}}_{,l m }({\vec p_2}) 
\Pi _{i' l',  j'm'}({\vec p_1})
\delta_{\vec{p}_1 + \vec{p}_2} 
. 
\end{eqnarray}
The final result reads
\begin{tcolorbox}[colframe=white,arc=0pt]
\be\label{CB2}
\langle \CB_{2, i j}(p_1)\CB_{2,  l m}(p_2)\rangle = 
 \frac{1}{N}\frac{1}{p_1^3}\Pi_{ij,lm}( \vec p_1)\delta_{\vec{p}_1+\vec{p}_2},
\ee
 \end{tcolorbox}
 \noindent
which coincides with the expression (\ref{real}). The calculation in momentum space is much more complex and lengthy than the one in real space. We have written it in some detail to familiarise the reader with general aspects of the $Q$-formalism, and provide the necessary tools required for 
the calculation of the spin-2  three-point correlator.

\section{The  three-point correlator of the spin-2 field}

The three-point function presents a more involved structure. We are going to follow the prescription highlighted in the previous sections and used in the computation of the two-point function working directly in momentum space.
The three-point function  of the spin-2 field  can be obtained computing the correlator of three dual fields as in Eq.~\eqref{Bexpan2} and shadow transforming the result. We start with
\begin{align}
\label{grav3}
\langle B_2(p_1|z_1)B_2(p_2|z_2)B_2(p_3|z_3)\rangle &=-\frac{16\sqrt{2}}{N^3}\int_{k_1}\int_{k_2}\int_{k_3}\,\bigl\langle \, :\!Q_{k_1}Q_{p_1-k_1}\!: \, :\!Q_{k_2}Q_{p_2-k_2}\!: \, :\!Q_{k_3}Q_{p_3-k_3}\!: \, \bigr\rangle \nonumber \\
&\times [(z_1\cdot k_1)^2+(z_1\cdot(p_1-k_1))^2+6(z_1\cdot k_1)z_1\cdot(k_1-p_1)] \nonumber \\
&\times [(z_2\cdot k_2)^2+(z_2\cdot(p_2-k_2))^2+6(z_2\cdot k_2)z_2\cdot(k_2-p_2)] \nonumber \\
&\times [(z_3\cdot k_3)^2+(z_3\cdot(p_3-k_3))^2+6(z_3\cdot k_3)z_3\cdot(k_3-p_3)].
\end{align}
Following similar steps to those performed in \eqref{wick2}, the first piece in the integral becomes
\begin{equation}\label{Qwick}
\bigl\langle \, :\!Q_{k_1}Q_{p_1-k_1}\!: \, :\!Q_{k_2}Q_{p_2-k_2}\!: \, :\!Q_{k_3}Q_{p_3-k_3}\!: \, \bigr\rangle=   \frac{2 N}{2 \, k_1^2 k_2^2 k_3^2}  \, \delta_{\vec{p}_1-\vec{k}_1+\vec{k}_2}\delta_{\vec{p}_2-\vec{k}_2+\vec{k}_3}\delta_{\vec{p}_3-\vec{k}_3+\vec{k}_1},
\end{equation}
where the $2N$ factor comes from the traces over the O$(2N)$ group indices.

In order to maintain the explicit symmetry under a cyclic permutation of $(\vec p_1, \vec p_2, \vec p_3)$, we symmetrise by adding $1/3$ times each of the three equivalent choices of deltas. The three choices are the following
\begin{eqnarray}
&\text{(a)}& \qquad \vec k_2 =\vec k_1 - \vec p_1\, , \qquad \vec k_3 = \vec k_1+\vec p_3~, \nonumber \\
&\text{(b)}& \qquad \vec  k_1 =\vec k_3 - \vec p_3\, , \qquad \vec k_2 =\vec  k_3+\vec p_2~, \nonumber \\
&\text{(c)}& \qquad \vec  k_1 =\vec  k_2 +\vec  p_1\, , \qquad \vec  k_3 =\vec   k_2 -\vec   p_2~.
\end{eqnarray}
We perform two integrals using the delta functions and reconstruct the external momentum conservation $\delta_{\vec p_1+\vec p_2+\vec p_3}$. After some manipulations we obtain
\begin{align}
\label{grav3.c}
\langle B_2(p_1|z_1)B_2(p_2|z_2)B_2(p_3|z_3)\rangle &=
-\frac{32\sqrt{2}}{3 N^2} \, \delta_{\vec{p}_1+\vec{p}_2+\vec{p}_3}
\int_k  \frac{1}{\vec k^2} \frac{1}{| \vec k - \vec p_1|^2} \frac{1}{| \vec k + \vec p_2|^2}
\nonumber \\
&\times [(z_1\cdot (k-p_1))^2+(z_1\cdot k)^2+6(z_1\cdot k)z_1\cdot(k-p_1)]
 \nonumber \\
&\times [(z_2\cdot k  )^2+(z_2\cdot(k + p_2))^2+6(z_2\cdot k)z_2\cdot(k + p_2)]
 \nonumber \\
&\times [(z_3\cdot (k + p_2))^2+(z_3\cdot \lp k- p_1 \rp)^2+6 z_3\cdot (k + p_2) \lp z_3\cdot \lp k- p_1 \rp  \rp ]
\nonumber \\
&+\text{all with }(p_1 \to p_3, p_2\to p_1, p_3\to p_2,z_1 \to z_3, z_2\to z_1, z_3\to z_2) 
\nn \\
&+\text{all with }(p_1 \to p_2, p_2\to p_3, p_3\to p_1,z_1 \to z_2, z_2\to z_3, z_3\to z_1).
\end{align}
After  fully expanding  this expression (containing 54 terms + cyclic permutations) , we can identify the following set of  independent integrals (the lower index identifies the number of $k$'s  in  the integrals)
\be
 \mathcal{I}_{r,q_1 q_2}^{i_1 \cdots i_r} &=
\int_k \frac{1}{\vec k^2} 
\frac{1}{| \vec k - \vec q_1|^2} \frac{1}{| \vec k + \vec q_2|^2} 
\lp 
 k^{i_1} \cdots  k^{i_r} \rp .
\ee
Notice that, by construction, the integrals $\I_{r,q_1 q_2}^{i_1 \cdots i_r}$ are symmetric under the exchange of any couple of indices.
Equation \eqref{grav3.c} then can be written as
\begin{align}
\label{BBB.e}
&\langle B_2(p_1|z_1)B_2(p_2|z_2)B_2(p_3|z_3)\rangle =
-\frac{32\sqrt{2}}{3N^2} \, \delta_{\vec{p}_1+\vec{p}_2+\vec{p}_3}
\bigg [
z_{1a}z_{1b} z_{2i}z_{2j}z_{3m}z_{3n}
 \Big (
512 \, \I_{6,p_1 p_2}^{abijmn} 
\nn\\
& -512  \, \I_{5,p_1 p_2}^{abijm} p_1^n +512  \, \I_{5,p_1 p_2}^{abijm} p_2^n+64  \, \I_{4,p_1 p_2}^{abij} p_1^m p_1^n
-384 \, \I_{4,p_1 p_2}^{abij}   p_1^m p_2^n+64 \, \I_{4,p_1 p_2}^{abij}  p_2^m p_2^n
\nn\\
&+512 \, \I_{5,p_1 p_2}^{abimn}   p_2^j
-512 \, \I_{4,p_1 p_2}^{abim} p_1^n p_2^j+512 \, \I_{4,p_1 p_2}^{abim} p_2^j p_2^n
+64 \, \I_{3,p_1 p_2}^{abi} p_1^m p_1^n p_2^j
\nn\\
&-384 \, \I_{3,p_1 p_2}^{abi}  p_1^m p_2^j p_2^n+64 \, \I_{3,p_1 p_2}^{abi}  p_2^j p_2^m p_2^n
+64\, \I_{4,p_1 p_2}^{abmn} p_2^i p_2^j -64 \, \I_{3,p_1 p_2}^{abm} p_1^n p_2^i p_2^j
\nn \\
  &+64 \, \I_{3,p_1 p_2}^{abm} p_2^i p_2^j  p_2^n
 +8 \, \I_{2,p_1 p_2}^{ab} p_1^m p_1^n p_2^i p_2^j
  -48 \, \I_{2,p_1 p_2}^{ab} p_1^m  p_2^i p_2^j  p_2^n+8\, \I_{2,p_1 p_2}^{ab} p_2^i p_2^j  p_2^m p_2^n
 \nn \\
  &-512 \, \I_{5,p_1 p_2}^{aijmn} p_1^b
  +512 \, \I_{4,p_1 p_2}^{aijm} p_1^b p_1^n
  -512 \, \I_{4,p_1 p_2}^{aijm} p_1^b p_2^n-64 \, \I_{3,p_1 p_2}^{aij} p_1^b p_1^m p_1^n
 \nn \\
  &+384 \, \I_{3,p_1 p_2}^{aij}  p_1^b p_1^m p_2^n-64 \, \I_{3,p_1 p_2}^{aij}  p_1^b p_2^m p_2^n
  -512 \, \I_{4,p_1 p_2}^{aimn} p_1^b p_2^j  +512 \, \I_{3,p_1 p_2}^{aim} p_1^b p_1^n p_2^j
 \nn \\
   &-512 \, \I_{3,p_1 p_2}^{aim} p_1^b p_2^j p_2^n-64 \, \I_{2,p_1 p_2}^{ai} p_1^b p_1^m p_1^n p_2^j
   +384 \, \I_{2,p_1 p_2}^{ai} p_1^b p_1^m p_2^j p_2^n
   -64 \, \I_{2,p_1 p_2}^{ai} p_1^b p_2^j p_2^m p_2^n
  \nn \\
   &-64  \, \I_{3,p_1 p_2}^{amn}   p_1^b p_2^i p_2^j +64  \, \I_{2,p_1 p_2}^{am} p_1^b p_1^n p_2^i p_2^j
    -64  \, \I_{2,p_1 p_2}^{am} p_1^b p_2^i p_2^j  p_2^n-8  \, \I_{1,p_1 p_2}^{a} p_1^b p_1^m p_1^n p_2^i p_2^j 
   \nn \\
     &   +48  \, \I_{1,p_1 p_2}^{a} p_1^b p_1^m p_2^i p_2^j  p_2^n
    -8  \, \I_{1,p_1 p_2}^{a} p_1^b p_2^i p_2^j p_2^m p_2^n
   +64 \, \I_{4,p_1 p_2}^{ijmn} p_1^a p_1^b-64 \, \I_{3,p_1 p_2}^{ijm} p_1^a p_1^b p_1^n
  \nn \\
   & +64 \, \I_{3,p_1 p_2}^{ijm} p_1^a p_1^b p_2^n
   +8 \, \I_{2,p_1 p_2}^{ij} p_1^a p_1^b p_1^m p_1^n
   -48 \, \I_{2,p_1 p_2}^{ij} p_1^a p_1^b p_1^m p_2^n+8 \, \I_{2,p_1 p_2}^{ij} p_1^a p_1^b p_2^m p_2^n   
 \nn  \\
   & +64 \, \I_{3,p_1 p_2}^{imn} p_1^a p_1^b p_2^j-64 \, \I_{2,p_1 p_2}^{im} p_1^a p_1^b p_1^n p_2^j
   +64 \, \I_{2,p_1 p_2}^{im} p_1^a p_1^b p_2^j p_2^n
   +8\, \I_{1,p_1 p_2}^{i} p_1^a p_1^b p_1^m p_1^n p_2^j
 \nn  \\
  &  -48 \, \I_{1,p_1 p_2}^{i} p_1^a p_1^b p_1^m p_2^j p_2^n+8\, \I_{1,p_1 p_2}^{i} p_1^a p_1^b p_2^j p_2^m p_2^n
   +8 \, \I_{2,p_1 p_2}^{mn} p_1^a p_1^b p_2^i p_2^j 
   -8 \, \I_{1,p_1 p_2}^{m} p_1^a p_1^b p_1^n p_2^i p_2^j
  \nn \\
   & +8 \, \I_{1,p_1 p_2}^{m} p_1^a p_1^b p_2^i p_2^j  p_2^n+ \, \I_{0,p_1 p_2} p_1^a p_1^b p_1^m p_1^n p_2^i p_2^j
     -6 \, \I_{0,p_1 p_2}  p_1^a p_1^b p_1^m p_2^i p_2^j  p_2^n+ \, \I_{0,p_1 p_2} p_1^a p_1^b p_2^i p_2^j  p_2^m p_2^n
     \Big)
   \nn  \\
&+\text{ all with }(p_1 \to p_3, p_2\to p_1, p_3\to p_2,z_1 \to z_3, z_2\to z_1, z_3\to z_2) 
 \nn\\
&+\text{ all with }(p_1 \to p_2, p_2\to p_3, p_3\to p_1,z_1 \to z_2, z_2\to z_3, z_3\to z_1)
\Big].
\end{align}
At this point we can get rid of the $z_i$ following the polynomial prescription as in Eq.~\eqref{polexpBBB}.

\subsection{The general structure}
First, let us consider the most general structure expected  to be found. The general form of a three-point function of  spin-2 fields is known in literature (see for example Ref.~\cite{Bzowski:2013sza})
and it is of the form
\be
\langle B_{2, ab}(p_1)B_{2, ij}(p_2)B_{3, mn}(p_3)\rangle = \langle t_{ ab}(p_1)t_{ ij}(p_2)t_{ mn}(p_3)\rangle + \text{ longitudinal and trace terms },
\ee
where $t_{ ab}(p_1) = {\Pi^{a' b'}}_{,ab}(\vec p_1) B_{2, a'b'}(p_1)$, identifying the transverse and traceless component of the spin-2 dual field.
Reconstructing the full expression can be done in principle, but it can be very hard in practice. Since we are interested in the three-point function  of spin-2 fields, we perform the shadow transform which requires a contraction with three projectors as in Eq.~\eqref{BCB}
\begin{equation}
\label{ShadowBBB}
\langle \CB_{2, ab}(p_1)\CB_{2, ij}(p_2)\CB_{3, mn}(p_3)\rangle =
 \frac{1}{p_1^3p_2^3p_3^3} 
 {\Pi^{a'b'}}_{,ab}({\vec p_1})
{ \Pi^{i' j'}}_{,ij}({\vec p_2})
{\Pi^{m' n'}}_{,mn}({\vec p_3}) 
\langle B_{2, a' b'}(p_1)B_{2, i' j'}(p_2)B_{2, m' n'}(p_3)\rangle.
\end{equation}
The projectors select the transverse and traceless part and we get
  \begin{tcolorbox}[colframe=white,arc=0pt]
\vspace{-.35cm}
\begin{equation}\label{BBB.ske}
\begin{aligned}
&\langle \CB_{2, ab}(p_1)\CB_{2, ij}(p_2)\CB_{3, mn}(p_3)\rangle =
\frac{ 1}{p_1^3p_2^3p_3^3}  \delta_{\vec p_1 + \vec p_2 +\vec p_3 }
\Pi_{ab,i_1i_2}({\vec p_1})\Pi_{ij, i_3 i_4}({\vec p_2})\Pi_{mn, i_5 i_6}({\vec p_3}) 
\\
&
\left [
A_1 p_2^{i_1} p_2^{i_2} p_3^{i_3} p_3^{i_4} p_1^{i_5} p_1^{i_6}  
\right .
\\
& + A_2 \delta^{i_2 i_4} p_2^{i_1} p_3^{i_3} p_1^{i_5} p_1^{i_6} + A_2(p_1 \leftrightarrow p_3) \delta^{i_4 i_6} p_2^{i_1} p_2^{i_2} p_3^{i_3} p_1^{i_5} +
 A_2(p_2 \leftrightarrow p_3) \delta^{i_2 i_6} p_2^{i_1} p_3^{i_3} p_3^{i_4} p_1^{i_5} 
\\
 &+ A_3 \delta^{i_1 i_3} \delta^{i_2 i_4} p_1^{i_5} p_1^{i_6} + A_3(p_1 \leftrightarrow p_3) \delta^{i_3 i_5} \delta^{i_4 i_6} p_2^{i_1} p_2^{i_2} + A_3(p_2 \leftrightarrow p_3) \delta^{i_1 i_5} \delta^{i_2 i_6} p_3^{i_3} p_3^{i_4}  
 \\
&\left . +A_4 \delta^{i_1 i_5} \delta^{i_3 i_6} p_2^{i_2} p_3^{i_4}
 + A_4(p_1 \leftrightarrow p_3) \delta^{i_1 i_5} \delta^{i_3 i_2} p_3^{i_4} p_1^{i_6} +
 A_4(p_2 \leftrightarrow p_3) \delta^{i_1 i_3} \delta^{i_5 i_4} p_2^{i_2} p_1^{i_6} \right .
 \\
& \left . + A_5 \delta^{i_1 i_4} \delta^{i_3 i_6} \delta^{i_5 i_2} 
  \right ].
  \end{aligned}
  \end{equation}
   \end{tcolorbox}
   \noindent
   Therefore, by applying the three projectors on $\langle B_{2, ab}(p_1)B_{2, ij}(p_2)B_{3, mn}(p_3)\rangle $ given by Eq.~\eqref{BBB.e}, one can identify the coefficients $A_i$. In practice, in Eq. \eqref{BBB.e}, we can neglect terms which contains $\delta_{a' b'}$, $\delta_{i' j'}$ and $\delta_{m' n'}$, given that they were already zero since $z_1^2=z_2^2=z_3^2=0$. Furthermore, we can neglect pieces which would give rise to contractions of the form
\be
\vec p_1 \cdot \Pi (\vec p_1)=0\,, \qquad \vec p_2 \cdot \Pi (\vec p_2)=0\,, \qquad \vec p_3 \cdot \Pi (\vec p_3)=0,
\ee
since the projector is transverse by construction.

\subsection{Result}
The symmetry properties of the correlator outlined in the previous subsection lead to a simplification of Eq.~\eqref{BBB.e}, which becomes
\be
\langle B_{2}^{a b}(p_1)B_{2}^{i j}(p_2)B_{3}^{m n}(p_3)\rangle &\supset
-\frac{32\sqrt{2}}{3N^2} \, \delta_{\vec{p}_1+\vec{p}_2+\vec{p}_3}
\bigg [
 64 p_1^{mn} \I_{4, p_1 p_2}^{a b i j}
 -384 p_1^{m} p_2^{n}     \I_{4, p_1 p_2}^{abij}
 \\
   &+64 p_2^{m n} \I_{4, p_1 p_2}^{ab i j}
   +64 p_1^{ij} \I_{4, p_1 p_3}^{a b m n}
   -384 p_1^{i} p_3^{j} \I_{4, p_1 p_3}^{a b m n}
   +64 p_3^{i j} \I_{4, p_1 p_3}^{a b m n}
   \\
 &   +64 p_2^{a b} \I_{4, p_2 p_3}^{i j m n}
   -384 p_2^{a} p_3^{b} \I_{4, p_2 p_3}^{i j m n}
   +64 p_3^{a b}  \I_{4, p_2 p_3}^{i j m n}
   -512 p_1^{n}    \I_{5, p_1 p_2}^{a b i j m}
   \\
   &+512 p_2^{n}  \I_{5, p_1 p_2}^{a b i j m}
   +512 p_1^{j}  \I_{5, p_1 p_3}^{a b i m n}
   -512 p_3^{j}  \I_{5, p_1 p_3}^{a b i m n}
   -512 p_2^{b}  \I_{5, p_2 p_3}^{a i j m n}
   \\
   &+512 p_3^{b} \I_{5, p_2 p_3}^{a i j m n}
   +512   \I_{6, p_1 p_2}^{a b i j m n}
   +512   \I_{6, p_1 p_3}^{a b i j m n}
   +512  \I_{6, p_2 p_3}^{a b i j m n}
\Big].
\ee
The previous expression is fully symmetric under a cyclic permutation of the external momenta as required by construction.
The structure in Eq. \eqref{BBB.ske} is fully recovered and, after a substitution of the integrals computed in Appendix \ref{app:rec}, the coefficients are found to be 

\begin{subequations}
\be
A_1&= -\frac{32\cdot 8192\sqrt{2}}{3N^2 } \big(
 i_{3,4,\{1,3,3\}}+ i_{3,4,\{3,1,3\}}+ i_{3,4,\{3,3,1\}}-4 \, i_{3,5,\{1,3,4\}}
\\
& \qquad  \qquad \qquad \ \  -4 \, i_{3,5,\{1,4,3\}}-4 \, i_{3,5,\{3,1,4\}}-4 \, i_{3,5,\{3,4,1\}}-4 \, i_{3,5,\{4,1,3\}}
\\
& \qquad  \qquad \qquad \ \ -4 \, i_{3,5,\{4,3,1\}}+4 \, i_{3,6,\{1,3,5\}}+8 \, i_{3,6,\{1,4,4\}}+4 \, i_{3,6,\{1,5,3\}}
\\
& \qquad  \qquad \qquad \ \ +4 \, i_{3,6,\{3,1,5\}}+4 \, i_{3,6,\{3,5,1\}}+8 \, i_{3,6,\{4,1,4\}}+8 \, i_{3,6,\{4,4,1\}}
\\
& \qquad  \qquad \qquad \ \ +4 \, i_{3,6,\{5,1,3\}}+4 \, i_{3,6,\{5,3,1\}} 
\big),
\ee
\be
A_{2}& =-\frac{32\cdot 8192\sqrt{2}}{3N^2 }  \big(
  i_{3,3,\{2,2,1\}}+2 \, i_{3,4,\{1,2,3\}}+2 \, i_{3,4,\{2,1,3\}}-4 \, i_{3,4,\{2,3,1\}}
\\
& \qquad  \qquad \qquad \ \ -4 \, i_{3,4,\{3,2,1\}}-4 \, i_{3,5,\{1,2,4\}}-4 \, i_{3,5,\{1,3,3\}}-4 \, i_{3,5,\{2,1,4\}}
\\
& \qquad  \qquad \qquad \ \ +4 \, i_{3,5,\{2,4,1\}}-4 \, i_{3,5,\{3,1,3\}}+8\, i_{3,5,\{3,3,1\}}+4 \, i_{3,5,\{4,2,1\}}
 \big),
\ee
\be
A_{3}&=-\frac{32\cdot 1024\sqrt{2}}{3N^2 } \big(
 \, i_{3,2,\{1,1,1\}}-4\, i_{3,3,\{1,2,1\}}-4 \, i_{3,3,\{2,1,1\}}+8 \, i_{3,4,\{1,1,3\}}
\\
& \qquad  \qquad \qquad \ \ +4 \, i_{3,4,\{1,3,1\}}+4 \, i_{3,4,\{2,2,1\}}+4 \, i_{3,4,\{3,1,1\}}
\big),
\ee
\be
A_{4}&=-\frac{32\cdot 8192\sqrt{2}}{3N^2 } \big(
 \, i_{3,3,\{1,2,1\}}+ \, i_{3,3,\{2,1,1\}}-2 \, i_{3,4,\{1,2,2\}}-2 \, i_{3,4,\{1,3,1\}}
\\
& \qquad  \qquad \qquad \ \ -2 \, i_{3,4,\{2,1,2\}}+2 \, i_{3,4,\{2,2,1\}}-2 \, i_{3,4,\{3,1,1\}}
\big),
\ee
\be
A_5 &=-\frac{32\cdot12288\sqrt{2}}{3N^2 }
 \, i_{3,3,\{1,1,1\}},
\ee
\end{subequations}
where $i_{3,m,\{n_1,n_2,n_3\}}$ are defined in Eq. (\ref{e:idmj}). 
The coefficients $A_1,A_5$ are symmetric under the cyclic permutation of the external momenta, while the coefficients $A_2,A_3,A_4$ are symmetric under the exchange $\vec p_1 \leftrightarrow \vec p_2$.
The integrals in $A_1,A_2$ are all convergent, so they can easily be computed using the formulas shown in Appendix\hspace{0.1cm}\ref{feynman}. However $A_3,A_4,A_5$ contain only divergent integrals for which we need to use the regularisation procedure as described in Appendix\hspace{0.1cm}\ref{app:reg}.

Our final result is 
  \begin{tcolorbox}[colframe=white,arc=0pt]
\vspace{-.35cm}
\begin{equation}
\begin{aligned}
\label{Acoeff}
A_1 &= C_{1} \frac{8}{a_{123}^6} \left[ a_{123}^3 + 3 a_{123} b_{123} + 15 c_{123} \right], 
\\
A_2 & = C_{1}  \frac{8 }{a_{123}^5} \left[ 4 p_3^4 + 20 p_3^3 a_{12} + 4 p_3^2 (7 a_{12}^2 + 6 b_{12}) + 15 p_3 a_{12} (a_{12}^2 + b_{12}) + 3 a_{12}^2 (a_{12}^2 + b_{12}) \right],
\\
A_3 & =C_{1}  \frac{2 p_3^2}{a_{123}^4} \left[ 7 p_3^3 + 28 p_3^2 a_{12} + 3 p_3 (11 a_{12}^2 + 6 b_{12}) + 12 a_{12} ( a_{12}^2 + b_{12} ) \right] 
\\
&
 - C_{2} \frac{8 \sqrt{\pi} }{3 a_{123}^2} \left[ a_{123}^3 - a_{123} b_{123} - c_{123} \right], 
\\
A_4 & = C_{1} \frac{4}{a_{123}^4} \left[ -3 p_3^5 - 12 p_3^4 a_{12} - 9 p_3^3 (a_{12}^2 + 2 b_{12}) + 9 p_3^2 a_{12} (a_{12}^2 - 3 b_{12}) \right.
\\
& \left. + \: (4 p_3 + a_{12}) (3 a_{12}^4 - 3 a_{12}^2 b_{12} + 4 b_{12}^2) \right] 
- \: C_{2} \frac{16 \sqrt{\pi} }{3 a_{123}^2} \left[ a_{123}^3 - a_{123} b_{123} - c_{123} \right],
\\
A_5 & = C_{1}  \frac{2}{a_{123}^3} \left[ -3 a_{123}^6 + 9 a_{123}^4 b_{123} + 12 a_{123}^2 b_{123}^2 - 33 a_{123}^3 c_{123} + 12 a_{123} b_{123} c_{123} + 8 c_{123}^2 \right] 
\\
& +  C_{2} \: \frac{8}{3} \sqrt{\pi}  (p_1^3 + p_2^3 + p_3^3),
\end{aligned}
\end{equation}
\end{tcolorbox}
\noindent
where 
\begin{align}
	C_1= -\frac{8 \sqrt{2}} { 15 N^2}, \qquad C_2= -\frac{6 \sqrt{2}}{ \sqrt{\pi} N^2},
\end{align} 
and we have used the following compact notation for the external momenta
\be
& a_{123} = p_1 + p_2 + p_3, \qquad b_{123} = p_1 p_2 + p_1 p_3 + p_2 p_3, \qquad c_{123} = p_1 p_2 p_3, \nn\\
& a_{ij} = p_i + p_j, \qquad\qquad \quad  b_{ij} = p_i p_j,
\ee
with $i,j = 1,2,3$. 
As a  sanity check for the coefficients of the three-point function of the spin-2 field, we have   controlled that  they correctly satisfy the primary conformal Ward identities as outlined in Ref.~\cite{Bzowski:2013sza}.

\section{The three-point correlator as Einstein plus (Weyl)$^{3}$}
We are now in the position to elaborate further about the  three-point correlator of the spin-2 field. In the introduction we reminded the reader that general symmetric arguments  impose  that the three-point correlator of the spin-2 field must be a combination of the three-point correlators induced by the Einstein and the (Weyl)$^{3}$ terms in the action. Given the complexity of the formulae (\ref{BBB.ske}) and (\ref{Acoeff}),  this looks miraculous, but it is indeed a beautiful example of symmetry in action.
 
We define the spin-2 field as
\begin{equation} \label{hsqf}
\CB_{2, i j}(\eta,\vx) =  \sum_\lambda  \int_k   \,\epsilon_{ij}^\lambda(\vk) \,\CB^{ \lambda}_{\vec {k}}(\eta)\,e^{i \vec {k} \cdot \vec x}.
\end{equation}
By shadow transforming the results of the previous section and using the
 property of the projector tensor in Eq.~\eqref{defproj}, 
 
 \begin{align}    
\epsilon^{* \lambda_1}_{ab}(\vec p_1) {\Pi^{ab}}_{,j_1j_2}({\vec p_1}) = \epsilon^{* \lambda_1}_{j_1 j_2}({\vec p_1}),
\end{align}
we obtain
\begin{align}
\Big< \CB^{ \lambda_1}_{\vec p_1 }\CB^{ \lambda_2}_{\vec p _2} \CB^{ \lambda_3}_{\vec p _3} \Big> = 
\epsilon^{* \lambda_1}_{ab}(\vec p_1)
 \epsilon^{* \lambda_2}_{ij} (\vec p_2)
  \epsilon^{* \lambda_3}_{mn} (\vec p_3)
  \langle \CB_{2}^{ ab}(p_1)\CB_{2}^{ij}(p_2)\CB_{3}^{mn}(p_3)\rangle.
\end{align}
Upon defining the
following quantities
\begin{eqnarray}\label{Est}
{\cal E}_1^\lambda(\vk_1| \vk_2,\vk_3)&=&\epsilon^{*\lambda}_{ij}(\vk_1)k_2^i k_3^j, \nonumber\\
{\cal E}_2^{\lambda_1\lambda_2}(\vk_1,\vk_2| \vk_3,\vk_4)&=&
\epsilon^{* \lambda_1}_{ik}(\vec k_1)
 \epsilon^{* \lambda_2,k}_{j} (\vec k_2)k_3^i k_4^j,\nonumber\\
{\cal E}_3^{\lambda_1\lambda_2}(\vk_1,\vk_2) &=& \epsilon^{* \lambda_1}_{ij}(\vec k_1)
 \epsilon^{* \lambda_2,ij}(\vec k_2),\nonumber\\
  {\cal E}_4^{\lambda_1\lambda_2\lambda_3}(\vk_1,\vk_2,\vk_3|\vk_4,\vk_5)&=&
\epsilon^{* \lambda_1}_{ij}(\vec k_1)
 \epsilon^{* \lambda_2}_{kl}(\vec k_2)
  \epsilon^{* \lambda_3,ik}(\vec k_3)k_4^j k_5^l,\nonumber\\
   {\cal E}_5^{\lambda_1\lambda_2\lambda_3}(\vk_1,\vk_2,\vk_3)&=&
\epsilon^{* \lambda_1}_{ij}(\vec k_1)
 \epsilon^{* \lambda_2,i}_{k}(\vec k_2)
   \epsilon^{* \lambda_3,jk}(\vec k_3),
\end{eqnarray}
we can rewrite the three-point correlator as 
\begin{eqnarray}
\label{aaa}
\Big< \CB^{ \lambda_1}_{\vec p_1 }\CB^{ \lambda_2}_{\vec p _2} \CB^{ \lambda_3}_{\vec p _3} \Big> &=&\frac{ 1}{p_1^3p_2^3p_3^3}  \delta_{\vec p_1 + \vec p_2 +\vec p_3 }\left[A_1 {\cal E}_1^{\lambda_1}(\vp_1| \vp_2,\vp_2){\cal E}_1^{\lambda_2}(\vp_2| \vp_3,\vp_3){\cal E}_1^{\lambda_3}(\vp_3| \vp_1,\vp_1)\right.\nonumber\\
&+& A_2 {\cal E}_2^{\lambda_1\lambda_2}(\vp_1,\vp_2| \vp_2,\vp_3) {\cal E}_1^{\lambda_3}(\vp_3| \vp_1,\vp_1)
+A_2(p_1 \leftrightarrow p_3){\cal E}_2^{\lambda_2\lambda_3}(\vp_2,\vp_3| \vp_3,\vp_1){\cal E}_1^{\lambda_1}(\vp_1| \vp_2,\vp_2)
\nonumber\\
&+&A_2(p_2 \leftrightarrow p_3){\cal E}_2^{\lambda_1\lambda_3}(\vp_1,\vp_3| \vp_2,\vp_1){\cal E}_1^{\lambda_2}(\vp_2| \vp_3,\vp_3)+ A_3 {\cal E}_3^{\lambda_1\lambda_2}(\vp_1,\vp_2){\cal E}_1^{\lambda_3}(\vp_3| \vp_1,\vp_1)\nonumber\\
&+&A_3(p_1 \leftrightarrow p_3)
 {\cal E}_3^{\lambda_2\lambda_3}(\vp_2,\vp_3){\cal E}_1^{\lambda_1}(\vp_1| \vp_2,\vp_2)
 +A_3(p_2 \leftrightarrow p_3)
 {\cal E}_3^{\lambda_1\lambda_3}(\vp_1,\vp_3){\cal E}_1^{\lambda_2}(\vp_2| \vp_3,\vp_3)\nonumber\\
 &+& A_4 {\cal E}_4^{\lambda_1\lambda_2\lambda_3}(\vp_1,\vp_2,\vp_3|\vp_2,\vp_3)+
 A_4(p_1 \leftrightarrow p_3) {\cal E}_4^{\lambda_3\lambda_2\lambda_1}(\vp_3,\vp_2,\vp_1|\vp_1,\vp_3)\nonumber\\
 &+&\left.A_4(p_2 \leftrightarrow p_3) {\cal E}_4^{\lambda_1\lambda_3\lambda_2}(\vp_1,\vp_3,\vp_2|\vp_2,\vp_1)
 +A_5 {\cal E}_5^{\lambda_1\lambda_2\lambda_3}(\vp_1,\vp_2,\vp_3)\right].
\end{eqnarray}
In Appendices  \ref{pol} and \ref{epsi} we provide the explicit expressions for these structures  in the $X$ and $P$ basis and in the chiral basis, respectively. 
We use here  the same notation of Ref. \cite{mp} for the $X$ and $P$ basis (usually dubbed the $\left\{\times,+\right\}$ basis).  

We note that the last  term  in the coefficient $A_5$ of Eq. (\ref{Acoeff})  (the piece proportional to $C_2$) is a contact term and it parametrises the ambiguity in the definition of the spin-2 field. Indeed,  it     can always be removed upon redefining the $\CB_{ij}$ field by $\CB_{ij}\rightarrow \CB_{ij}+c\,\CB_{ik}\CB^{k}_j$ ($c$ being a constant), see for instance the discussion in Ref. \cite{mp}.\footnote{Notice that the contact terms are necessary to reproduce the exact squeezed   limit of the graviton three-point functions from the stress tensor three-point function \cite{soft1,soft2}. We thank P. McFadden and K. Skenderis for discussions about this point.} We disregard it from now on.

We define the quantities
\begin{eqnarray}
J(p_1,p_2,p_3)&=&2(p_1^2p_2^2+p_1^2p_3^2+p_2^2p_3^2)-(p_1^4+p_2^4+p_3^4), \nonumber \\
I(p_1,p_2,p_3)&=& \left(p_1+p_2+p_3-\frac{(p_1p_2+p_1 p_3+p_2 p_3)}{p_1+p_2+p_3}-
\frac{p_1 p_2 p_3}{(p_1+p_2+p_3)^2}\right),
\end{eqnarray}
to construct the following shapes  
\begin{eqnarray}
{\rm E}^{PPP}(p_1,p_2,p_3)&=&\frac{J(p_1,p_2,p_3)}{4(p_1 p_2 p_3)^5}\left(\sum_{i=1}^3p_i^4+6\sum_{i<j}p_i^2 p_j^2\right) I(p_1,p_2,p_3), \nonumber\\
{\rm E}^{XXP}(p_1,p_2,p_3)&=&\frac{J(p_1,p_2,p_3)}{(p_1 p_2 p_3)^4}\frac{p_1^2+p_2^2+3 p_3^2}{p_3} I(p_1,p_2,p_3), \nonumber \\
W^{3\, PPP}(p_1,p_2,p_3)&=&270\frac{(p_1+p_2-p_3)(p_2+p_3-p_1)(p_3+p_1-p_2)}{(p_1+p_2+p_3)^3(p_1 p_2 p_3)^2},\nonumber\\
W^{3\, XXP}(p_1,p_2,p_3) &=&-W^{3\, PPP}(p_1,p_2,p_3).
\end{eqnarray}
They represent the shapes  from the Einstein term and the (Weyl)$^{3}$ term of the three-point correlators \cite{mp}.
Using the tensorial contractions in Appendix\hspace{0.1cm}\ref{pol}, we can now write the long expression for the three-point correlators from the higher-spin gravity in a simple and
compact form\footnote{This simplification can be also understood as originated by the tensorial degeneracies which exist in three dimensions, see for instance Ref. \cite{Bzowski:2013sza}.}

\begin{tcolorbox}[colframe=white,arc=0pt]
\vspace{-.15cm}
\begin{equation}
\begin{aligned}
\label{ew}
\Big< \CB^{ P}_{\vec p_1 }\CB^{ P}_{\vec p _2} \CB^{ P}_{\vec p _3} \Big>' &= -\frac{2}{N^2}{\rm E}^{PPP}(p_1,p_2,p_3)+\frac{8}{270N^2}W^{3\, PPP}(p_1,p_2,p_3), 
\\
\Big< \CB^{ X}_{\vec p_1 }\CB^{ X}_{\vec p _2} \CB^{ P}_{\vec p _3} \Big> '&= \frac{2}{N^2}{\rm E}^{XXP}(p_1,p_2,p_3)+\frac{8}{270 N^2} W^{3\, XXP}(p_1,p_2,p_3),
\end{aligned}
\end{equation}
\end{tcolorbox}
\noindent
where we have used the standard notation for which the $\langle\cdots\rangle'$ indicates that the factor $(2\pi)^3$ times the Dirac delta has been removed. 

\begin{figure}[t]
	\includegraphics[width=0.49 \linewidth]{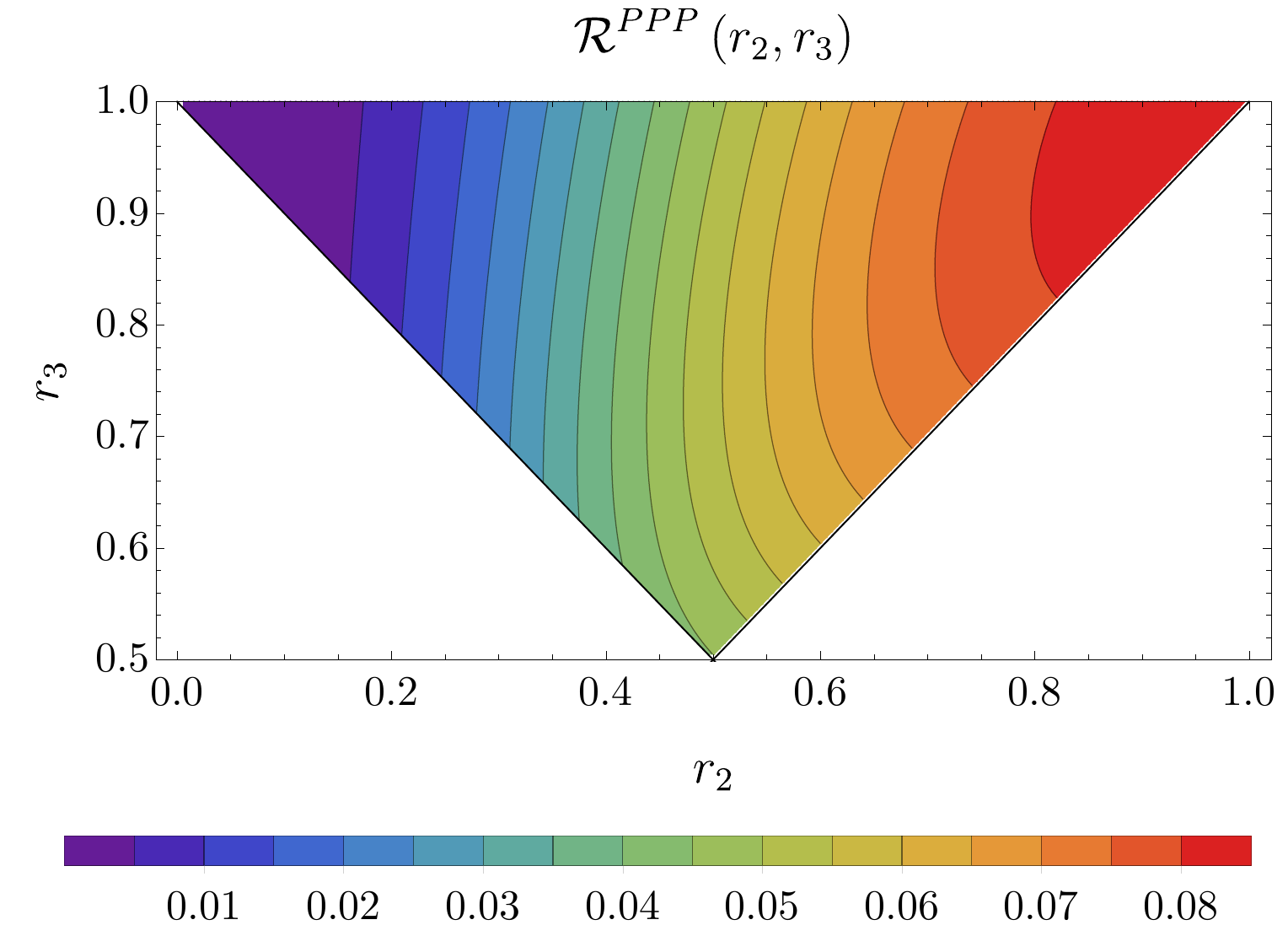}
		\includegraphics[width=0.49 \linewidth]{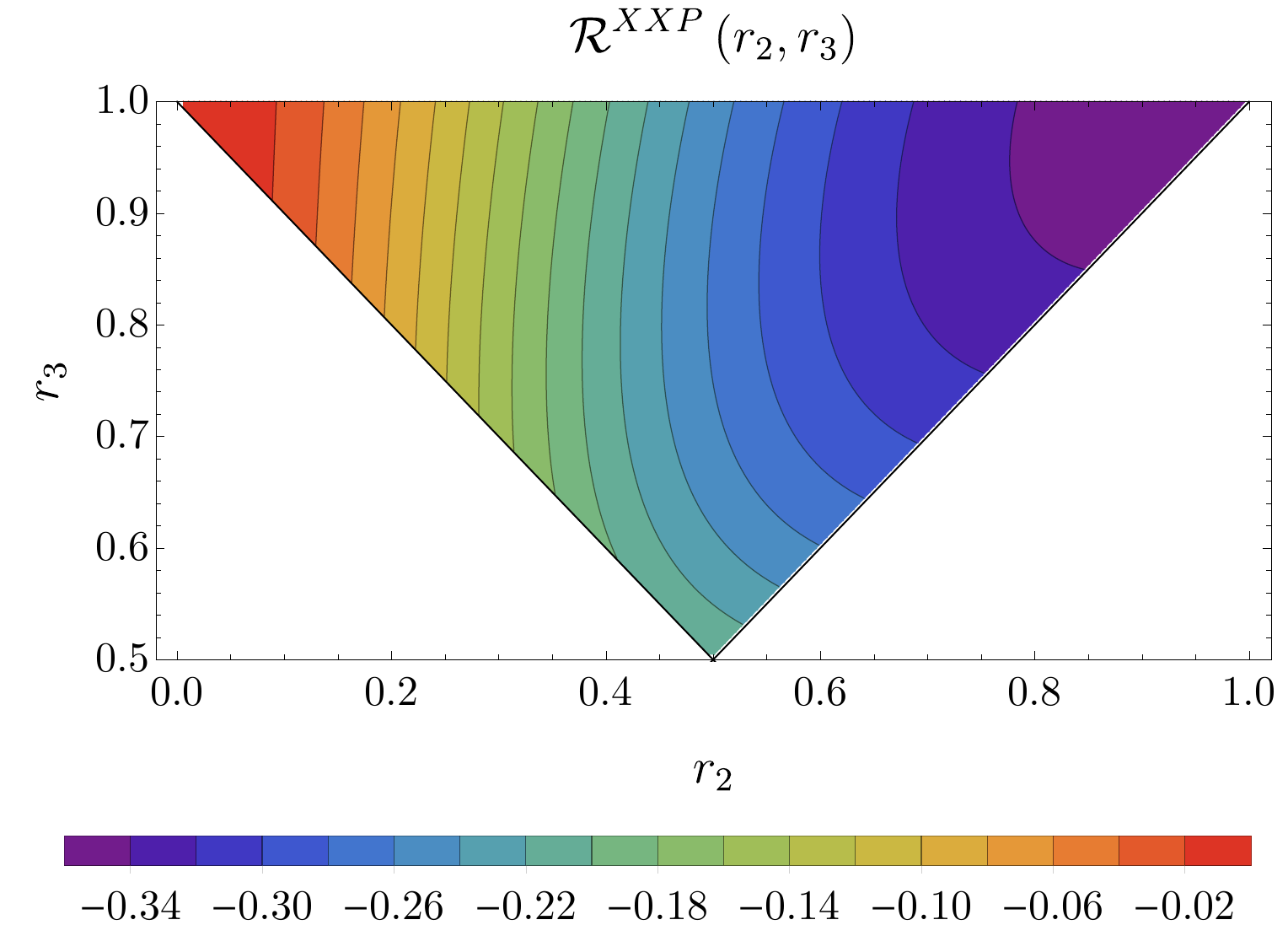}
	\centering
	\caption{
	Ratios $\mathcal{R}^{\lambda_1,\lambda_2,\lambda_3}(r_2,r_3)$ as defined in Eq.~\eqref{ratios}.}
	\label{fig:shape1}
\end{figure}

Thus, we have established that the three-point correlator for the spin-2 field in Vasiliev's higher-spin minimal bosonic  theory are a combination of two pieces, one coming from the Einstein term and the other from the (Weyl)$^{3}$ term, once the contact terms have been properly removed. Notice also that, even though shape-dependent,  the (Weyl)$^{3}$ is parametrically of the same order of the Einstein term. Using the gravitational action (\ref{einstein}), one finds that the ratio of the (Weyl)$^{3}$-to-the-Einstein contributions is (up to the shapes)
equal to $(HL)^4/2$ \cite{mp}.  From our results we infer that 
\be
\label{hl}
HL= \left(\frac{8}{270}\right)^{1/4}\simeq 0.4.
\ee
The scale $L$   gives a measure of the size of higher derivative corrections and in  higher-spin gravity this length scale turns out to be rather sizeable. 
In Fig.~\ref{fig:shape1} we plot the ratio of the contribution from the   (Weyl)$^{3}$ and the  Einstein term, as a function of the ratios $r_2= p_2/p_1$, $r_3= p_3/p_1$
\be\label{ratios}
\mathcal{R}^{PPP}(r_2,r_3)=\frac{W^{3\, PPP}(p_1,p_2,p_3)}{{\rm E}^{ PPP}(p_1,p_2,p_3)}
\,\,\,\,{\rm and}\,\,\,\,
\mathcal{R}^{XXP}(r_2,r_3)=\frac{W^{3\, XXP}(p_1,p_2,p_3)}{{\rm E}^{ XXP}(p_1,p_2,p_3)}.
\ee
The previous quantities are independent of $p_1$. 
We see that this ratio is maximised for the so-called equilateral configuration, where all the momenta are equal. This makes sense as the (Weyl)$^3$ term depends on gradients of the spin-2 field.
\noindent

We  dedicate  the next section to a more refined discussion of the shapes of the three-point correlator for the spin-2 field.

\section{The shapes of the graviton three-point correlator} 
In this section we wish to analyse the possible shape configurations of the graviton three-point correlator. In order to make contact with the majority of the literature, we do so in the chiral basis, see Appendix\hspace{0.1cm}\ref{epsi}, where
the helicity R   corresponds to $\lambda=1$ and the helicity ${\rm L}$ corresponds to $\lambda=-1$.  
We may consider two different limits. 

\subsection{The squeezed limit}
In the squeezed limit we take $p_1\ll p_2,p_3$ and  obtain
\be\label{BBB.squeezed}
\begin{aligned}
	\Big< \CB^{ \lambda_1}_{\vec p_1 }\CB^{ \lambda_2}_{\vec p _2} \CB^{ \lambda_3}_{\vec p _3} \Big> '_{p_1 \ll p_2,p_3}  &= 
	24 \sqrt{2} \Big< \CB^{ \lambda_1}_{\vec {p}_1}\CB^{ \lambda_1}_{-\vec {p}_1}\Big>'\Big< \CB^{ \lambda_2}_{\vec {p}_2}\CB^{ \lambda_3}_{\vec {p}_3}\Big>'
\frac{1}{p_2^2}{\cal E}_1^{\lambda_1}(\vp_1| \vp_2,\vp_2).
\end{aligned} 
\ee
At this stage we can  use the argument that conformal symmetries fix the squeezed limit of  the three-point correlator to match the one obtained in Ref. \cite{maldacenaNG}.
We have the freedom to
normalise the spin-2 field as 
\be
\CB_{ij} =\frac{1}{8\sqrt{2}}  \gamma_{ij},
\ee
where $\gamma_{ij}$ is the graviton field. Its two-point correlator is 
\be
\langle  \gamma_\vk^\lambda   \gamma_{-\vk}^{\lambda'}  \rangle '= \frac{1}{128 N} \frac{\delta ^{\lambda \lambda'}}{ k^3},
\ee
and by choosing $N=M_{\rm p}^2/128H^2$  one obtains the correct normalisation.
We then obtain
\begin{tcolorbox}[colframe=white,arc=0pt]
\vspace{-.25cm}
\begin{equation}\label{BBB.squeezed}
\begin{aligned}
	\Big< \gamma^{ \lambda_1}_{\vec p_1 }\gamma^{ \lambda_2}_{\vec p _2} \gamma^{ \lambda_3}_{\vec p _3} \Big>'_{\vp_1\ll \vp_2,\vp_3}  &= 
	\frac{ 3}{p_1^3 p_2^3}
 \delta^{\lambda_2 \lambda_3}
\epsilon^{* \lambda_1}_{i_1 i_2}(\vec p_1) \frac{p_2^{i_1} p_2^{i_2} }{p_2^2},
\end{aligned}	
\end{equation}
\end{tcolorbox}
\noindent
which is the standard squeezed limit result  \cite{maldacenaNG} after accounting for notational differences.  
This does not come as a surprise. The  tensor consistency relation  in the squeezed limit is very robust and basically  equivalent to the adiabaticity of the tensor
perturbations on super-Hubble scales. Adiabaticity is a property of the long wavelength tensor modes in higher-spin gravity too as one can easily understand inspecting, for instance, the (Weyl)$^3$ graviton interactions which depend only on time derivatives of the tensor fields. Therefore the tensor consistency relation is respected in higher-spin gravity. 
A confirmation of this result comes also from the  scalar-scalar-graviton three-point function calculated in Ref. \cite{anninos} 

\begin{align}
\label{002ev}
\langle \CB_{0}(p_1)\CB_{0}(p_2)\CB_{2, ij}(p_3)\rangle 
=  \frac{{16}\sqrt{2}}{N^2}\frac{1}{p_3^3} \frac{p_1 + p_2 + 2p_3}{(p_1 + p_2 + p_3)^2}   \delta_{\vec{p}_1+\vec{p}_2+\vec{p}_3} \Pi_{ij,i'j'}({\vec p_3})  
p_1^{i'}p_2^{j'}.
\end{align}
By indicating with 
 $\phi_{\vec p}$ the scalar with mass $2H^2$, the squeezed limit of such a three-point function reads
\begin{tcolorbox}[colframe=white,arc=0pt]
\vspace{-.35cm}
\begin{align}
\label{002squeezed}
\Big< \gamma^{ \lambda}_{\vec p_1 }\phi_{\vec p _2} \phi_{\vec p _3} \Big>'_{p_1\ll p_2, p_3} 
= 
\Big< \gamma^{ \lambda'}_{\vec {p}_1}\gamma^{ \lambda'}_{-\vec {p}_1}\Big>'
\Big< \phi_{\vec {p}_2}\phi_{\vec {p}_3}\Big>' 
\epsilon^{* \lambda}_{i_1 i_2}(\vec p_1) p_2^{i_1} p_2^{i_2}
,
\end{align}
\end{tcolorbox}
\noindent
which is indeed the expected squeezed limit taking into account that the conformal weight of the scalar field is a conformal primary field of dimension $\Delta=1$.

\subsection{The equilateral limit}
The equilateral limit $p \equiv p_1=p_2=p_3$ may be also immediately obtained  by inserting in the expression (\ref{aaa}) the coefficients (in $A_5$ we have removed the contact
term)
\begin{eqnarray}
A_1&=&-\frac{1472\sqrt{2}}{3645 N^2 p^3},\,\,\,\,A_2=-\frac{1664\sqrt{2}}{243 N^2 p},\,\,\,\,
A_3=\frac{3488\sqrt{2}p}{135 N^2}, \,\,\,\,
A_4=\frac{22592\sqrt{2}p}{405 N^2}, \,\,\,\,
A_5=-\frac{3152\sqrt{2}p^3}{405 N^2},\nonumber\\
\end{eqnarray}
and evaluating the polarisation structures shown in Appendix\hspace{0.1cm}\ref{epsi}, in such a limit.
The result is
\vskip 0.2cm
  \begin{tcolorbox}[colframe=white,arc=0pt]
\vspace{-.35cm}
\be
\hspace{-0.35cm}
\Big< \gamma^{ \lambda_1}_{\vec p_1 }\gamma^{ \lambda_2}_{\vec p_2 } \gamma^{ \lambda_3}_{\vec p_3 } \Big>'_{p_1 = p_2 = p_3} = 
	\frac{-3197 - 3044(\lambda_1 \lambda_2 +\lambda_2 \lambda_3 + \lambda_1 
\lambda_3)}{5184} \lp \Big< \gamma^{ \lambda_1}_{\vec {p}}\gamma^{ \lambda_1}_{-\vec {p}}\Big>'
	\Big< \gamma^{ \lambda_2}_{\vec {p}}\gamma^{ \lambda_2}_{-\vec {p}}\Big>' +{\rm perm.}\rp .
\ee\end{tcolorbox}

\subsection{The generic shape}
\begin{figure}[t]
	\includegraphics[width=0.49 \linewidth]{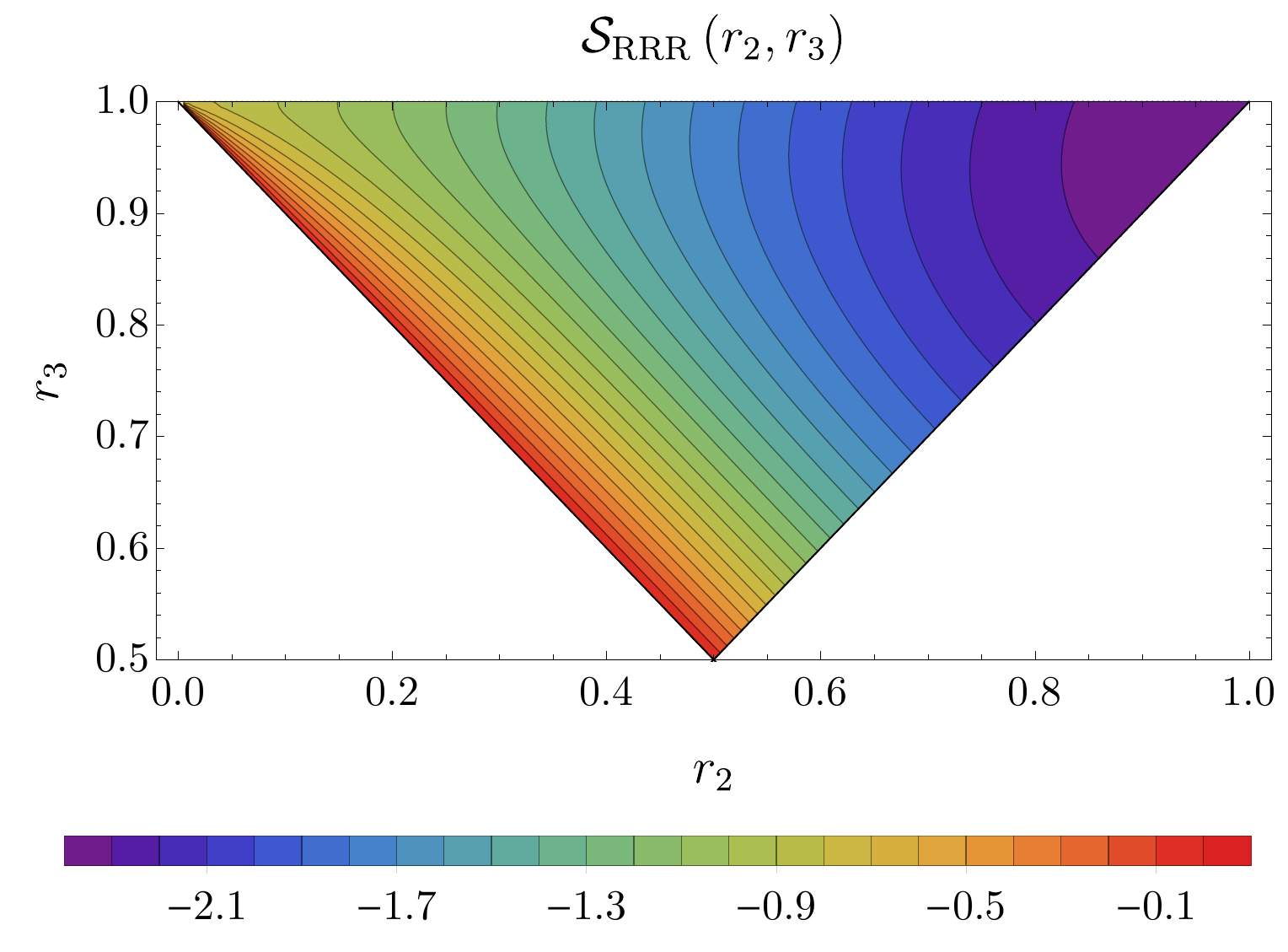}
		\includegraphics[width=0.49 \linewidth]{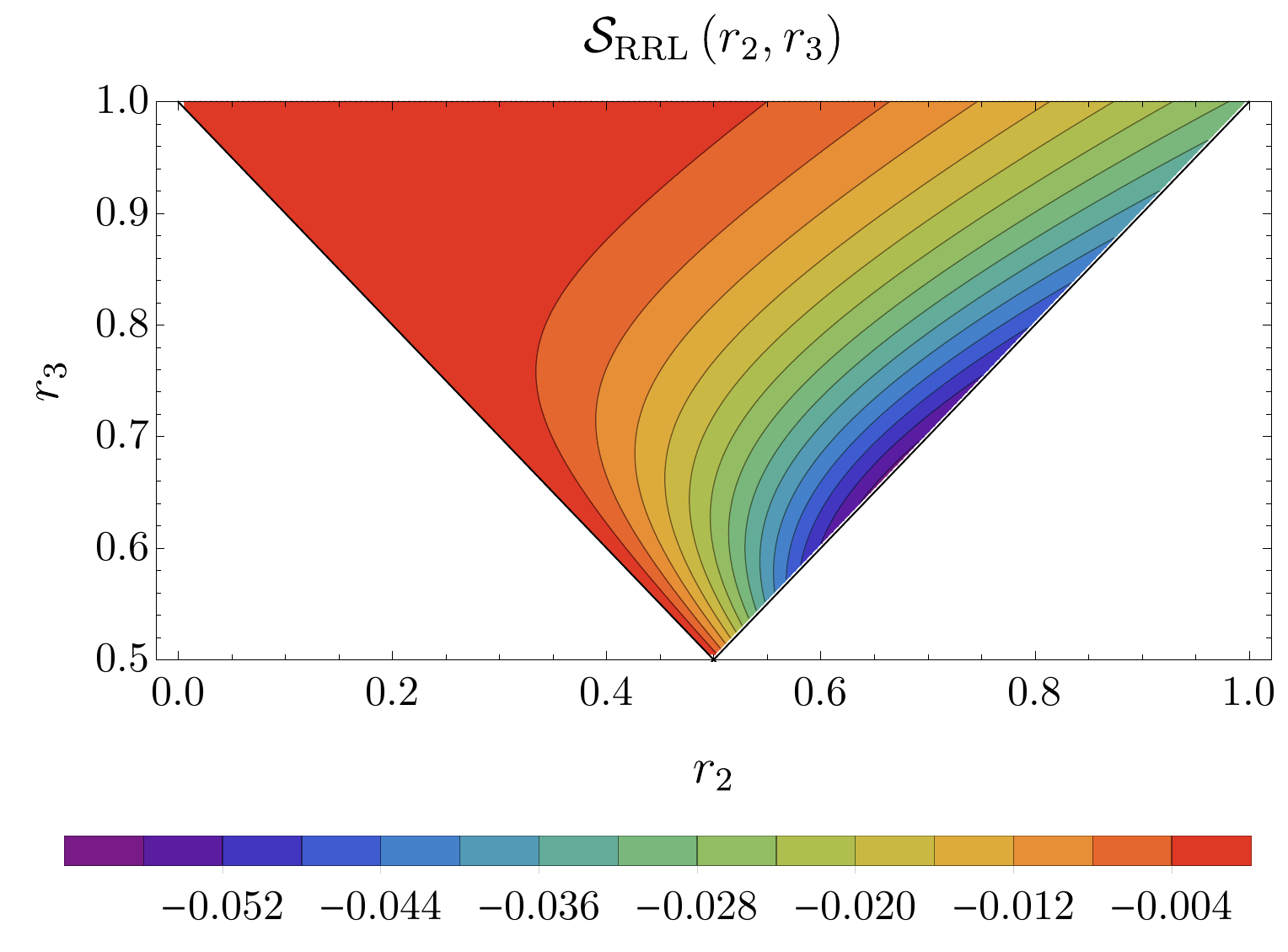}
			\includegraphics[width=0.49 \linewidth]{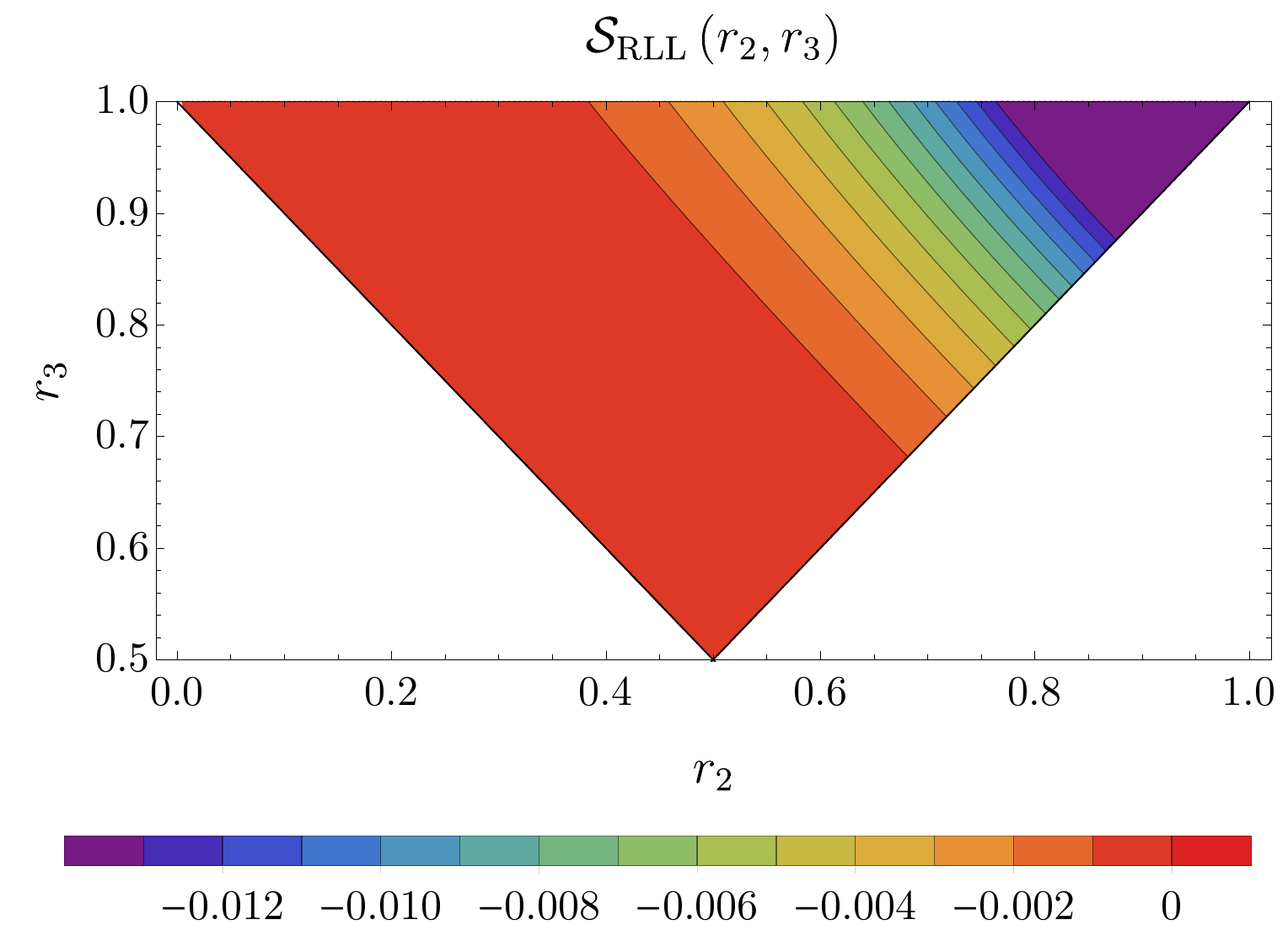}
		\includegraphics[width=0.49 \linewidth]{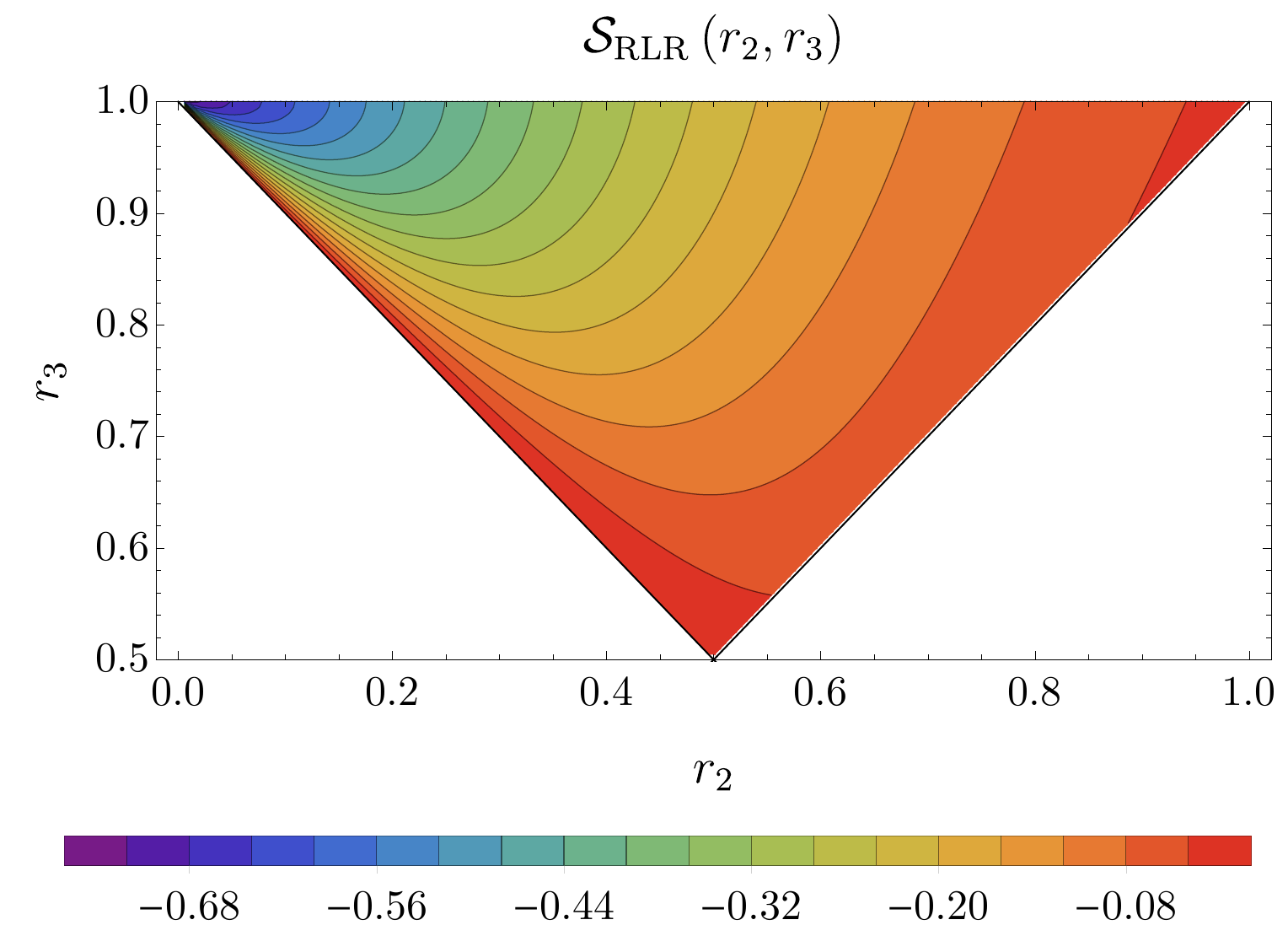}
	\centering
	\caption{Shapes $\mathcal{S}_{\lambda_1,\lambda_2,\lambda_3}(r_2,r_3)$ for the independent combinations of the polarisations.}
	\label{fig:shape}
\end{figure}

We can identify a few  standard shapes for the  graviton three-point correlators:  the  local one where the signal is dominated by squeezed configuration $p_1\ll  p_2 \simeq  p_3$; the equilateral  configuration whose signal is enhanced at the configuration  $p_1\simeq p_2\simeq p_3$ ; the  folded configuration whose signal is maximised  at   $p_1 + p_2 \simeq p_3$; and finally the  orthogonal configuration ($p_1\simeq p_2$) generating a   signal with a positive enhancement  at the equilateral configuration and a negative peak in  the folded configuration. 
In Fig.~\ref{fig:shape} we plot the shape defined as
\begin{equation}
	\mathcal{S}_{\lambda_1,\lambda_2,\lambda_3}(r_2,r_3)=
	\frac{\Big< \gamma^{ \lambda_1}_{\vec {p}_1}\gamma^{\lambda_2}_{\vec {p}_2}\gamma^{\lambda_3}_{\vp_3}\Big>'}{  
	\Big< \gamma^{ \lambda_1}_{\vec {p}_1}\gamma^{ \lambda_1}_{-\vec {p}_1}\Big>'
		\Big< \gamma^{ \lambda_2}_{\vec {p}_2}\gamma^{ \lambda_2}_{-\vec {p}_2}\Big>'
	+ {\rm perm.} },\,\,\,\, r_2=\frac{p_2}{p_1}\,\,\,\,{\rm and}\,\,\,\,  r_3=\frac{p_3}{p_1},
\end{equation}
for several combinations of the polarisations. The largest signal comes from the RRR combination (and therefore for the LLL combination) and is maximised for the equilateral configuration.
\noindent
\hspace{-0.15cm}

It is not clear what the  observational  prospects are for measuring  the shapes  of the  tensor non-Gaussianities  in the  fortunate case that primordial gravitational waves from inflation are detected. At first sight, detecting tensor three-point correlators  might look futuristic. A  quadrupolar anisotropy in the tensor power spectrum can be  induced by the non-Gaussian graviton three-point  peaked in the squeezed limit. However, we have seen that higher-spin gravity does not lead to an enhancement of such an anisotropy
with respect to the standard single-field model case. A study of the role played by the (Weyl)$^3$ term in the CMB temperature intensity as well as the  B-mode polarisation bispectra
can be found in Ref. \cite{sm} from which one can preliminary estimate\footnote{We thank M. Shiraishi for discussions about this point.}
  that a detection is possible if $L\gsim 10^{-5} M_{\rm p}^{-1}$. 

The first step towards characterising  tensor  non-Gaussianities with the Laser Interferometer Space Antenna (LISA)  has been very recently taken  in Ref. \cite{ng1} (see also Refs. \cite{ng2,ng3,ng4}) with  determination of the  interferometer three-point response functions. Even though a detailed study is not the scope of our paper, it would be certainly interesting to further investigate the detectability of the tensor non-Gaussianities from higher-spin gravity, with particular attention to the presence of the (Weyl)$^3$-induced terms.

\section{Conclusions}

In this paper we have investigated non-Gaussian features of the graviton in the minimal Vasiliev theory. The theory contains an infinite tower of higher-spin fields for each even spin, and admits asymptotically de Sitter configurations.
By exploiting the formalism of \cite{anninos}, we have calculated the exact three-point correlation function of the massless spin-2 field at late times. In accordance with symmetry arguments, we have  shown that the shape of the graviton correlators is  a linear combination of the shape produced by the standard Einstein term and the shape produced by a term cubic in the Weyl tensor. The Vasiliev model fixes the relative weight of these two pieces. We conclude with a few comments.

While have considered the possibility of a higher-spin phase during inflation, we have not provided a mechanism to exit from the higher-spin phase into the current universe. Any such mechanism, as well as the physics responsible for the observed Gaussian scalar fluctuations, will go beyond the Vasiliev framework. It might be interesting to consider a curvaton-like scenario  to generate the scalar fluctuations with the observed scalar tilt \cite{c1,c2,c3} whereby during the de Sitter epoch the role of the curvaton might be played  by some additional light scalar field. 
Perhaps the problem of Higgsing the higher-spin gauge symmetry \cite{higgshs,Chang:2012kt} down to the diffeomorphism group is relevant to these questions. 
We have also disregarded the correlators of higher-spin fields. The fate of such correlators is not clear (at least to us) once the universe has exited the de Sitter phase.\footnote{FRW-like solutions for higher-spin gravity have been constructed in Ref. \cite{frwhs}.} On the other hand, since the graviton correlators we have computed  are at super-Hubble wavelengths, they cannot be affected by local processes.

To provide evidence for our scenario, measuring the tensorial non-Gaussian  shapes, and in particular those sourced by the cubic Weyl term,  is of paramount importance.  Interestingly, a considerable value for $HL$ may provide evidence for additional degrees of freedom during inflation. In flat spacetime, higher-derivative cubic interactions of  the  massless graviton cause  causality problems \cite{cam}. The introduction of an infinite  tower of higher-spin states cures such a  drawback. Even though it is not known that the causality problem persists in a de Sitter spacetime, it is reasonable to speculate that  measuring a large higher-derivative graviton non-Gaussianity would indicate  the presence of a tower of higher-spin states. Along this vein, it would be interesting to compare and contrast the predictions of higher-spin gravity to those of the weakly coupled holographic models put forward in Refs.~\cite{hm1,hm2},  or other models of Vasiliev gravity with de Sitter solutions \cite{Anninos:2014hia,Chang:2013afa,Hertog:2017ymy}, where the  tensorial non-Gaussianities may also differ from those predicted by pure  Einstein gravity.  

As a final note, it is interesting to reflect on the picture painted in Ref. \cite{Strominger:2001gp} (see also \cite{Larsen:2003pf,Anninos:2012qw,Kiritsis:2013gia,hm1}). There, it is imagined that cosmological evolution between the current de Sitter era and that during inflation should be viewed, holographically, as a renormalisation group flow between two fixed-points. The current de Sitter era corresponds to the utraviolet fixed-point, and is strongly coupled. The inflationary era corresponds to the infrared fixed-point. If the infrared fixed point is moreover weakly coupled, it will contain an infinite tower of (almost) conserved currents. A bulk dual to a weakly coupled fixed point will contain a tower of light higher-spin particles \cite{Sundborg:2000wp,witten,Klebanov:2002ja,Sezgin:2002rt}. Consequently, our scenario corresponds to a flow from strong to weak coupling. While such a flow is certainly allowed, the full space renormalisation group flows is sufficiently rich and varied to prevent us from deeming our scenario natural in the absence of experiment. 

\acknowledgments
We gratefully acknowledge A.~Bzowski, F.~Denef, R.~Monten, M. Shiraishi, Z.~Sun, and especially P. McFadden and K. Skenderis for very useful correspondence and clarifications. D.A. is partially funded by the Royal Society under the grant {\sl The Atoms of a deSitter Universe}.
A.R. and G.F. are  supported by the Swiss National Science Foundation (SNSF), project {\sl The Non-Gaussian Universe and Cosmological Symmetries}, project number: 200020-178787.

\appendix

\section{Polarisations and projection tensors }
\label{projector}
\renewcommand\theequation{\Alph{section}.\arabic{equation}}
Polarisation tensors of higher-spin fields can  be obtained generalising the notion of polarisation vectors introducing positive and negative energy wave functions \cite{h6}
\begin{eqnarray}
\epsilon^\lambda_{i_1\cdots i_s}(\vk)&=&\sum_{\lambda_1,\cdots,\lambda_s=\pm 1}\delta_{\lambda_1+\cdots+\lambda_s,\lambda}
\sqrt{\frac{2^s(s+\lambda)!(s-\lambda)!}{(2s)!\prod_{i=1}^s(1+\lambda_i)!(1-\lambda_i)!}}\prod_{j=1}^s\epsilon^{\lambda_j}_{i_j}(\vk),\nonumber\\
\epsilon^{*\lambda}_{i_1\cdots i_s}(\vk)&=&\sum_{\lambda_1,\cdots,\lambda_s=\pm 1}\delta_{\lambda_1+\cdots+\lambda_s,\lambda}
\sqrt{\frac{2^s(s+\lambda)!(s-\lambda)!}{(2s)!\prod_{i=1}^s(1+\lambda_i)!(1-\lambda_i)!}}\prod_{j=1}^s\epsilon^{*\lambda_j}_{i_j}(\vk),
\end{eqnarray}
where $\epsilon^{\lambda}_{i}$ and $\epsilon^{*\lambda}_{i}$ are positive and negative energy wave functions for a spin-one field,
with
\be
\epsilon^{*\lambda}_{i}=(-1)^\lambda\epsilon^{-\lambda}_{i}.
\ee
It is useful to define the  projector tensor in $d$-dimensions as
\be\label{defproj}
\Pi_{ i_1 \cdots i_s, j_1 \cdots j_s} (\vk)
\equiv \sum_{\lambda} \epsilon^\lambda _{i_1 \cdots i_s} (\vk) \epsilon_{ j_1 \cdots j_s}^{*\lambda } (\vk). 
\ee
For instance, for spin-2 in three-dimensions we obtain (summing only over the maximally transverse modes as lower helicity states are zero in the transverse traceless gauge)
\be
\Pi_{i_1i_2, j_1j_2}(\vk) =\frac{1}{2}  \left(\Pi_{i_1, j_1}(\vk)  \Pi_{i_2,j_2}(\vk) + \Pi_{i_1,j_2} (\vk) \Pi_{i_2,j_1}(\vk)\right) - \frac{1}{2}\Pi_{i_1, i_2}(\vk)  \Pi_{j_1, j_2}(\vk).
\ee
Written explicitly in terms of the expanded spin-one projector $\Pi_{i,j}(\vk)= \delta_{ij} - \hat k_i \hat k_j$, one gets
\begin{align}
\Pi_{i_1i_2, j_1j_2}(\vk) &=
\frac{1}{2} \left [
\lp \delta_{i_1 j_1} \delta_{i_2 j_2} +\delta_{i_1 j_2} \delta_{i_2 j_1}-\delta_{i_1 i_2} \delta_{j_1 j_2} \rp 
+
 \hat k_{i_1} \hat k_{i_2} \hat k_{j_1} \hat k_{j_2}
-
\lp 
 \delta_{i_1 j_1}\hat k_{i_2} \hat k_{j_2}
 \right . \right . \nn
\\
& \left .\left .
+ \delta_{i_1 j_2}\hat k_{i_2} \hat k_{j_1}
+ \delta_{i_2 j_1}\hat k_{i_1} \hat k_{j_2} + \delta_{i_2 j_2}\hat k_{i_1} \hat k_{j_1}
- \delta_{i_1 i_2}\hat k_{j_1} \hat k_{j_2} - \delta_{j_1 j_2}\hat k_{i_1} \hat k_{i_2}
\rp 
\right].
\end{align}

\section{Shadow transform}
\label{shadow}
The shadow transform can be defined as follows.
From a primary field $O_{s,\Delta}(\vx)$ of spin-$s$ and scaling dimension $\Delta$ under the $d$-dimensional Euclidean conformal group SO$(1,d+1)$, we can construct a dual primary field 
\begin{align} \label{generalshadowtransf}
 \widetilde{O}_{s,  \widetilde \Delta}(\vx)=\int \dd^dy \, G_{s, \widetilde \Delta}(\vx-\vy) \, O_{s,\Delta}(\vy),
\end{align}
named shadow transform of $O_{s,\Delta}$, characterised by the same spin-$s$ and conjugate scaling dimension 
\begin{align}
 \widetilde  \Delta = d- \Delta \, .
\end{align}
The kernel of the transformation $G_{s, \widetilde  \Delta}(\vx-\vy)$ is represented by the two-point function of spin-$s$, dimension $ \widetilde  \Delta$ operators in a $d$-dimensional conformal field theory. For example, the shadow transform of a scalar field $O$ is
\begin{equation} \label{scalarshadowtr}
    \widetilde{O}_{\widetilde  \Delta}(\vx)=
    \int \dd^d y \, \frac{c_{\widetilde  \Delta}}{|\vx-\vy|^{2 \widetilde \Delta}} \, O_{\Delta}(\vy) \, 
\end{equation}
where $c_{\widetilde  \Delta}$ is a constant normalisation factor. 
Such a kernel makes  the operator $\widetilde O _{\widetilde  \Delta} (\vx)$ transform as a local primary field of dimension $\widetilde \Delta$ under the conformal group. It is also clear that the inverse of a shadow transform is again a shadow transform.

In momentum space, the shadow transform is simply obtained by Fourier transforming the real space result. 
For higher-spin fields, this procedure is more complicated due to the presence of tensor operators; however, as outlined in Ref.~\cite{anninos}, it is still possible to shadow relate the dual boundary fields $B_{s}(k)$ of conformal weight $\widetilde \Delta = s+1$ and the local ones $\CB_{s}(k)$ of conformal weight ${\Delta} = 2-s$ as 
\be
B_{s}^{i_1...i_s}(k) = G_{s,\widetilde \Delta=s+1}^{ i_1...i_s,m_1...m_s}(\vk, \vk ') \CB_{s, m_1...m_s}(k'),
\ee
where the Fourier transform of the shadow kernel is given, in the $d\rightarrow 3$ limit and for $s\geq 1$, by
\be
\label{FourierGreen}
G_{s, \widetilde \Delta=s+1}^{ i_1...i_s,m_1...m_s}(\vk, \vk ')
=\frac{(-1)^s\pi^2}{(2s)!}
c_{s, \widetilde \Delta=s+1}k^{2s-1}\Pi^{i_1...i_s,m_1...m_s}(\vk) \delta_{\vk +\vk'},
\ee
where $\Pi_{i_1...i_s,m_1...m_s}(\vk)$ is the projector presented previously in Appendix\hspace{0.1cm}\ref{projector}. For example, the shadow transformation rules for the spin-0 and spin-2 boundary fields is shown to be
\begin{equation}\label{BCB}
B_0(p)=\frac{1}{p}\CB_0(p) \, , \quad B_{ij}( p)=p^3 \, {\Pi^{i'j'}}_{,ij}({\vec p}) \,\CB_{2, i'j'}(p).
\end{equation}

\section{Double-K integrals}
\label{doublek}
In this appendix we show how to treat integrals of the form
\begin{equation}
\mathcal{I}_0^{i j l m}=
\int \frac{d^d \vec{k}}{(2\pi)^d} \frac{1}{|\vec{k}|^{2 \delta_1}} \frac{1}{|\vec{k} - \vec{p}_1|^{2 \delta_2}}
\lp k^i k^j k^l k^m \rp, 
\end{equation}
which we refer to as ``double-K integrals''. Integrals of this type 
enter in the computation of the graviton 2-point function.
Using the known Schwinger parametrization
\begin{equation}\label{Schwinger}
\frac{1}{A^\alpha} = \frac{1}{\Gamma(\alpha)}\int_{0}^{\infty}{\rm d}s \  s^{\alpha - 1}e^{-sA},
\end{equation}
with $\alpha > 0$, 
we find
\be
\mathcal{I}_0^{i j l m}=  \int \frac{{\rm d}^d k}{\lp 2 \pi \rp ^d}  k^i k^j k^l k^m
\int_{\mathbb{R}_+^2} {\rm d}s_1{\rm d}s_2
\ \frac{s_1^{\delta_1-1}}{\Gamma(\delta_1)}\frac{s_2^{\delta_2-1}}{\Gamma(\delta_2)} 
e^{-(s_1|\vec{k}|^2 + s_2|\vec{k}-\vec{p}_1|^2)}.
\ee
Next, we set $s_t=s_1+s_2$, $\vec l = \vec k - \frac{s_2 }{s_1+s_2} \vec p_1$ and $\Delta = \frac{s_1 s_2 }{s_t } p_1 ^2$, such that the integral can be written as
\begin{equation}
\int_{\mathbb{R}_+^2} {\rm d}s_1{\rm d}s_2\frac{s_1^{\delta_1-1}}{\Gamma(\delta_1)}\frac{s_2^{\delta_2-1}}{\Gamma(\delta_2)}  e^{- \Delta}
\int \frac{{\rm d}^d l}{\lp 2 \pi \rp ^d} e^{- s_t l^2}  \left[
\lp  l + \frac{s_2 }{s_1+s_2}  p_1 \rp^i
 \cdots
\lp  l + \frac{s_2 }{s_1+s_2}  p_1 \rp^m \right].
\end{equation}
This expression can be expanded and split up into a sum of integrals.
For any $a$ such that $a > 0$, the integral over $l$ can be performed as follows
\begin{align}
\label{int}
\int \frac{\d^d \vec{l}}{(2 \pi)^d} l^{2n} e^{-a l^2} & =
 \frac{\Gamma \left( \frac{d}{2} + n \right)}{(4 \pi)^{\frac{d}{2}} \Gamma \left( \frac{d}{2} \right)}
\frac{1}{a^{\frac{d}{2} + n}}, 
\\
\int \frac{\d^d \vec{l}}{(2 \pi)^d} l^{i_1} \ldots l^{i_{2n}} e^{-a l^2} & = \frac{S^{i_1 \ldots i_{2n}}}{(4 \pi)^{\frac{d}{2}} 2^n a^{\frac{d}{2} + n}}, 
\end{align}
while the presence of an odd number of $l$ makes the integrals vanish. We also defined  $S^{i_1 \ldots i_{2m}}$ to be a completely symmetric tensor with unitary coefficients as, for example,
\be
S^{i_1 i_2} &= \delta^{i_1 i_2}, 
\\
S^{i_1 i_2 i_3 i_4} &= \delta^{i_1 i_2} \delta^{i_3 i_4} + \delta^{i_1 i_3} \delta^{i_2 i_4} + \delta^{i_1 i_4} \delta^{i_2 i_3},
\\
{\rm S}^{i_1 i_2 i_3 i_4 i_5 i_6}
&= 
\delta ^{i_1 i_2}\delta ^{i_3 i_4} \delta^{i_5 i_6}+
\delta ^{i_1 i_2}\delta ^{i_3 i_5} \delta^{i_4 i_6}+
\delta ^{i_1 i_2}\delta ^{i_3 i_6} \delta^{i_4 i_5}
+
\delta ^{i_1 i_3}\delta ^{i_2 i_4} \delta^{i_5 i_6}
\\
&+
\delta ^{i_1 i_3}\delta ^{i_2 i_5} \delta^{i_4 i_6}+
\delta ^{i_1 i_3}\delta ^{i_2 i_6} \delta^{i_4 i_5}
+
\delta ^{i_1 i_4}\delta ^{i_2 i_3} \delta^{i_5 i_6}+
\delta ^{i_1 i_4}\delta ^{i_2 i_5} \delta^{i_3 i_6}
\\
&+
\delta ^{i_1 i_4}\delta ^{i_2 i_6} \delta^{i_3 i_5}
+
\delta ^{i_1 i_5}\delta ^{i_2 i_3} \delta^{i_4 i_6}+
\delta ^{i_1 i_5}\delta ^{i_2 i_4} \delta^{i_3 i_6}+
\delta ^{i_1 i_5}\delta ^{i_2 i_6} \delta^{i_3 i_4}
\\
&+
\delta ^{i_1 i_6}\delta ^{i_2 i_3} \delta^{i_4 i_5}+
\delta ^{i_1 i_6}\delta ^{i_2 i_4} \delta^{i_3 i_5}+
\delta ^{i_1 i_6}\delta ^{i_2 i_5} \delta^{i_3 i_4}.
\ee
Of all these pieces in the $l$ integral, let us focus on the one containing $l^i l^j  l^l l^m$ in the numerator, since it is the one proportional to the Dirac deltas. This is the one which is relevant for our computations, the others can be computed accordingly. Thus, we find
\begin{equation}
\int_{\mathbb{R}_+^2} {\rm d}s_1{\rm d}s_2
\frac{s_1^{\delta_1-1}s_2^{\delta_2-1}}{\Gamma^2}
e^{- \Delta}
\int \frac{{\rm d}^d l}{\lp 2 \pi \rp ^d} e^{- s_t l^2}  
 l ^i l^j l^l l^m=
 \frac{{\rm S}^{i j l m} }{4 \lp  4 \pi \rp  ^{\frac{d}{2}} 
 \Gamma^2}\int_{\mathbb{R}_+^2} {\rm d}s_1{\rm d}s_2 s_1^{\delta_1-1} s_2^{\delta_2-1}  s_t ^{- \lp \frac{d}{2} +2 \rp} e^{- \Delta},
\end{equation}
where $\Gamma^2=\Gamma(\delta_1) \Gamma(\delta_2)$.
We can change the variables $s_1,s_2$ to $v_1$ and $v_2$ as
\be 
s_1= \frac{\lp v_1 + v_2 \rp ^2}{2 v_1}=\frac{V}{2v_1} ,
\quad s_2= \frac{\lp v_1 + v_2 \rp ^2}{2 v_2}=\frac{V}{2v_2},
 \quad
 s_t= \frac{\lp v_1 +v_2  \rp^3}{2 v_1 v_2} =\frac{V^{3/2}}{2 v_1 v_2}.
\ee
The determinant of the Jacobian of the transformation is given by
\be
{\rm det} \left | J \right |=\left |   -\frac{\lp v_1 +v_2 \rp ^4} {4 v_1^2 v_2 ^2}\right | =\frac{V^2}{4 v_1^2 v_2^2},
\ee
such that the integral becomes
\be 
\mathcal{I}_0^{i j l m} \supset
 \frac{{\rm S}^{i j l m} }{ \lp  4 \pi \rp  ^{\frac{d}{2}} 2^{-\frac{d}{2} + \delta_1 +\delta_2} \Gamma(\delta_1) \Gamma(\delta_2)}
\int_{\mathbb{R}_+^2} {\rm d}v_1{\rm d}v_2 
\lp v_1 + v_2 \rp ^{2\lp  \delta_1 + \delta_2 -3 \frac{d}{4} -3\rp}
v_1^{\frac{d}{2} -\delta_1+1} v_2^{\frac{d}{2} -\delta_2+1}
 e^{- \lp \frac{v_1 +v_2}{2}\rp p_1^2}.
\ee
We introduce a new Schwinger parameter with $\alpha = 3d/2 +6 -2 \delta _t $ and $A= \lp v_1 +v_2 \rp/ \lp v_1 v_2 \rp  $, so that
\be 
\mathcal{I}_0^{i j l m} \supset
 \frac{2^{\frac{d}{2} - \delta_t}  {\rm S}^{i j l m} }{ \lp  4 \pi \rp  ^{\frac{d}{2}}  \Gamma(\delta_1) \Gamma(\delta_2) \Gamma \lp \frac{3d}{2} +6 -2 \delta _t \rp }
 \int_{\mathbb{R}_+^2} {\rm d}v_1{\rm d}v_2 \int_0 ^\infty {\rm d} t \,  t^{\frac{3d}{2} +5 -2 \delta _t} \prod _{j=1}^2
 e^{-\frac{t}{v_j}} e^{-\frac{v_j}{2} p_j^2}
 v_j ^{-d+2\delta _t -\delta_j -5}.
\ee
Now we perform the change of variables $u_i = p_1^2 v_i /2 $ and by using the definition of the Bessel functions 
\begin{equation}
K_j(z) = \frac{1}{2} \left( \frac{z}{2} \right)^j \int_0^\infty e^{-u-\frac{z^2}{4u}} u^{-j-1} \d u, \qquad | \arg z | < \frac{\pi}{4},
\end{equation}
we can  rewrite the integral as
\be 
\mathcal{I}_0^{i j l m} \supset
\frac{2^{\frac{-3d}{2} +2 \delta_t -8} \,  {\rm S}^{i j l m} }{ \lp  4 \pi \rp  ^{\frac{d}{2}}  \Gamma(\delta_1) \Gamma(\delta_2) \Gamma \lp \frac{3d}{2} +6 -2 \delta _t \rp 2^{-5 +\delta_t -\frac{1}{2}d}}\int_0 ^\infty {\rm d} t\,
 \lp \sqrt{2 t} \rp ^{d+2 -\delta_t} 
\\
\times 
 \prod_{i=1}^{2}
 p_i^{d -2\delta _t +\delta _i +4}K_{d-2 \delta _t +\delta _i +4}(\sqrt{2t} p_i),
\ee
where we introduced an additional vector $\vec p_2= - \vec p_1$ as usual in the 2-point function.
We redefine now the integration variable $x= \sqrt{2t}$ and get
\be
\mathcal{I}_0^{i j l m}
\supset \frac{2^{{-d} + \delta_t -3} \,  {\rm S}^{i j l m} }{ \lp  4 \pi \rp  ^{\frac{d}{2}}  \Gamma(\delta_1) \Gamma(\delta_2) \Gamma \lp \frac{3d}{2} +6 -2 \delta _t \rp } 
 \int_0 ^\infty {\rm d} x\,
x ^{d+3 -\delta_t}
 \prod_{i=1}^{2}
 p_i^{d -2\delta _t +\delta _i +4}K_{d-2 \delta _t +\delta _i +4}(x p_i).
\ee
Finally, defining the double-K integral as
\be\label{dkdef}
I_{\alpha\{\beta_1, \beta_2\}} (p_1,p_2)= \int_0^\infty \d x \: x^{\alpha } \prod_{j=1}^2 p_j^{\beta_j} K_{\beta_j}(p_j x),
\ee
we can write
\be
\mathcal{I}_0^{i j l m}
\supset \frac{2^{{-d} + \delta_t -3} \,  {\rm S}^{i j l m} }{ \lp  4 \pi \rp  ^{\frac{d}{2}}  \Gamma(\delta_1) \Gamma(\delta_2) \Gamma \lp \frac{3d}{2} +6 -2 \delta _t \rp } I_{d+3-\delta_t\{d-2 \delta _t +\delta _1 +4, d-2 \delta _t +\delta _2 +4\}} (p_1,p_2).
\ee

\section{Triple-K integrals}
\label{feynman}
In the computation of three-point functions like that of Eq.~\eqref{BBB.e}  we could incur in momentum space integrals of the form
\begin{equation}
\label{e:kint}
\I_{r, p_1 p_2} ^{i_1 \cdots i_r}  = \int \frac{d^d \vec{k}}{(2\pi)^d} \frac{k^{i_1} \cdots k^{i_r}}{|\vec{k}|^{2\delta_3}|\vec{k}-\vec{p}_1|^{2\delta_2}|\vec{k}+\vec{p}_2|^{2\delta_1}},
\end{equation}
where $\vec{p}_1,\vec{p}_2,\vec{p}_3$ identify three external momenta satisfing the condition $\vec{p}_1+\vec{p}_2+\vec{p}_3 = 0$. To solve such an integral we will follow the procedure outlined in Appendix A.3 of Ref.~\cite{Bzowski:2013sza}.
Using the  Schwinger parametrization defined in Eq.~\eqref{Schwinger} we can write the previous integral as
\begin{equation}\label{eqc3}
\I_{r, p_1 p_2} ^{i_1 \cdots i_r}  =\int \frac{d^d \vec{k}}{(2\pi)^d} k^{i_1} \cdots k^{i_r}\int_{\mathbb{R}_+^3} {\rm d}\vec{s} \ \frac{s_1^{\delta_1-1}}{\Gamma(\delta_1)}\frac{s_2^{\delta_2-1}}{\Gamma(\delta_2)} \frac{s_3^{\delta_3-1}}{\Gamma(\delta_3)} e^{-(s_3|\vec{k}|^2 + s_2|\vec{k}-\vec{p}_1|^2+ s_1|\vec{k}+\vec{p}_2|^2)},
\end{equation}
where ${\rm d}\vec{s} = {\rm d}s_1\d s_2 \d s_3$ and $\Gamma^3 = \Gamma(\delta_1)\Gamma(\delta_2)\Gamma(\delta_3)$. 
Then we can set $s_t = s_1 + s_2 + s_3$, $\vec{l} = \vec{k} + \frac{s_1\vec{p}_2 - s_2\vec{p}_1}{s_t}$ and $\Delta = \frac{s_1s_2p_3^2 + s_1s_3p_2^2 + s_2s_3p_1^2}{s_t}$, such that $s_3|\vec{k}|^2 + s_2|\vec{k}-\vec{p}_1|^2+ s_1|\vec{k}+\vec{p}_2|^2= s_t l^2 + \Delta$.
The integral in Eq.~\eqref{e:kint} then acquires the form
\begin{equation} \label{e:kint2}
\I_{r, p_1 p_2} ^{i_1 \cdots i_r}  =\Gamma^{-3} \int_{\mathbb{R}_+^3} \d \vec{s} \: s_1^{\delta_{1} - 1} s_2^{\delta_{2} - 1} s_3^{\delta_{3} - 1} e^{- \Delta} \int \frac{\d^d \vec{l}}{(2 \pi)^d} e^{-s_t l^2} \prod_{j=1}^r \left( l^{i_j} + \frac{s_2 p_1^{i_j} - s_1 p_2^{i_j}}{s_t} \right).
\end{equation}
This expression can be expanded and the integral in $l$ can be computed using Eq.~\eqref{int}. The result can be split up into a sum of integrals of the form
\begin{equation} \label{e:idmj}
i_{d,m,\{\delta_j\}} = \frac{1}{(4 \pi)^{\frac{d}{2}} 2^m \Gamma^3} \int_{\mathbb{R}_+^3} \d \vec{s} \: s_t^{- \frac{d}{2} - m} s_1^{\delta_{1} - 1} s_2^{\delta_{2} - 1} s_3^{\delta_{3} - 1} e^{- \Delta}.
\end{equation}
Be careful that the parameters $\delta_i$ are not necessarily the ones present in Eq.~\eqref{e:kint} because additional powers of $s_i$ can appear from the numerator product. On  the other hand, the numerical value of $\Gamma ^3$ is left unchanged.
Similarly to the computation performed in the previous section, we recast Eq.~\eqref{e:idmj} as:
\begin{align}
\label{e:Idm}
i_{d,m,\{\delta_j\}} 
& = \frac{2^{-\frac{d}{2}-2m+4}}{(4 \pi)^{\frac{d}{2}} \Gamma^3 \Gamma(d + 2m - \delta_t)} \int_0^\infty \d x \: x^{\frac{d}{2} + m - 1} \prod_{j=1}^3 p_j^{\frac{d}{2}+m-\delta_t+\delta_j} K_{\frac{d}{2}+m-\delta_t+\delta_j}(p_j x) \nonumber \\
& = \frac{2^{-\frac{d}{2}-2m+4}}{(4 \pi)^{\frac{d}{2}} \Gamma^3 \Gamma(d + 2m - \delta_t)} I_{\frac{d}{2} + m - 1 \{ \frac{d}{2} + m - \delta_t + \delta_j \}},
\end{align}
where $\delta_t = \delta_1 + \delta_2 + \delta_3$.
Finally, we isolated the structure of the triple-K integral
\be\label{tKi}
I_{\alpha\{\beta_1, \beta_2, \beta_3\}} (p_1,p_2,p_3)= \int_0^\infty \d x \: x^{\alpha } \prod_{j=1}^3 p_j^{\beta_j} K_{\beta_j}(p_j x),
\ee
which is going to be the building block for the solution of the considered integrals. 
Using the Bessel-K function identities
\begin{align}\label{Kprop}
\frac{\partial}{\partial a} \left[ a^\nu K_\nu(a x) \right] & = - x a^\nu K_{\nu - 1}(a x), \nn \\
K_{\nu-1}(x) + \frac{2 j}{x} K_{\nu}(x) & = K_{\nu + 1}(x), \nn \\
K_{-\nu}(x) & = K_{\nu}(x),
\end{align}
we find the following relations involving triple-K integrals
\begin{align}
\frac{\partial}{\partial p_n} I_{\alpha \{ \beta_j \}} & = - p_n I_{\alpha + 1 \{ \beta_j - \delta_{jn} \}}, \label{e:Jid1} \\
I_{\alpha \{ \beta_j + \delta_{jn} \}} & = p_n^2 I_{\alpha \{ \beta_j - \delta_{jn} \}} + 2 \beta_n I_{\alpha - 1 \{ \beta_j \}}, \label{e:Jid2} \\
I_{\alpha \{\beta_1 \beta_2, - \beta_3\}} & = p_3^{-2 \beta_3} I_{\alpha \{\beta_1 \beta_2 \beta_3\}}, \label{e:Jid4}
\end{align}
for any $n=1,2,3$. One can also arrange them to get the iterative formula
\begin{align}
\label{iterative div}
I_{\alpha + 1 \{ \beta_j + \delta_{jn} \}} = -p_n \frac{\partial}{\partial p_n} I_{\alpha \{ \beta_j \}} + 2 \beta_n I_{\alpha \{ \beta_j \}}.
\end{align}
One is typically able to handle these expressions in terms of simple polynomials of the external momenta, apart from when they are divergent and need to be regularised. 
This task is described in detail in the following appendix.

\subsection{Recursive formula}\label{app:rec}
Our computation involves many different integrals of the kind shown in \eqref{e:kint} and it is, therefore, useful to write down their relation to the triple-K integrals in terms of recursive formulas. In this section we are going to specialise the result to the case where $\delta_i$ defining the integral in Eq.~\eqref{e:kint} are equal to 1.
The structure of the integrals comes from the following product, see Eq.~\eqref{e:kint2}
\be
\I_{r, p_1 p_2} ^{i_1 \cdots i_r} 
=
 \int_{\mathbb{R}_+^3} \d \vec{s} \: s_1^{\delta_{1} - 1} s_2^{\delta_{2} - 1} s_3^{\delta_{3} - 1} e^{- \Delta}
	 \int \frac{\d^d \vec{l}}{(2 \pi)^d} e^{-s_t l^2}  \left [
	\lp l +\frac{ s_2 p_1 -s_1 p_2}{s_t}\rp^{i_1}  \cdots  	\lp l +\frac{ s_2 p_1 -s_1 p_2}{s_t}\rp^{i_r} \right].
	\ee 
The solution can be written in terms of a recursive formula,  once the expression of the polynomial $\prod_n (l+(s_2p_1-s_1 p_2)/s_t)^{i_n}$ has been expanded, as
\be
\I_{r, p_1 p_2} ^{i_1 \cdots i_r} =
 \sum_{\rm all \ terms } 
 \left [ 
(-1)^{n_{p_2}} 2^{n_{p_1} +n_{p_2}} 
\lp  i_{3,\frac{n_l}{2}+n_{p_1} +n_{p_2} ,\{ 1 +n_{p_2},1 +n_{p_1},1\}} \rp 
  p_1^{i _1 \cdots i_{n_{p_1}}} 
  p_2^{i _{n_{p_1}+1} \cdots i_{n_{p_1}+n_{p_2}}}
  \right ],
\ee 
where $n_l$ and $n_{p_i}$ stand for the number of $l$ and $p_i$, respectively, in the numerator of the considered piece. Notice that they manifest an explicit symmetry under the exchange of $p_1$ and $p_2$.
The integrals with cyclic permutations of $p_i$ are exactly the same once one has swapped the positions of the indices in $i_{d,m,\{\delta_i\}}$ accordingly. Thus we find
	\be
\I_{r, p_3 p_1} ^{i_1 \cdots i_r} =
 \sum_{\rm all \ terms } 
 \left [ 
(-1)^{n_{p_1}} 2^{n_{p_3} +n_{p_1}} 
\lp  i_{3,\frac{n_l}{2}+n_{p_3} +n_{p_1} ,\{1 +n_{p_3}, 1 , 1 +n_{p_1}\}} \rp 
  p_3^{i _1 \cdots i_{n_{p_3}}} 
  p_1^{i _{n_{p_3}+1} \cdots i_{n_{p_3}+n_{p_1}}}
  \right ],
  \\
  \I_{r, p_2 p_3} ^{i_1 \cdots i_r} =
 \sum_{\rm all \ terms } 
 \left [ 
(-1)^{n_{p_3}} 2^{n_{p_2} +n_{p_3}} 
 \lp i_{3,\frac{n_l}{2}+n_{p_2} +n_{p_3}, \{ 1, 1 +n_{p_3},1 +n_{p_2}\}} \rp 
  p_2^{i _1 \cdots i_{n_{p_2}}} 
  p_3^{i _{n_{p_2}+1} \cdots i_{n_{p_2}+n_{p_3}}}
  \right ].
\ee	
In the following we provide some examples of integrals involving the external momenta $\vec p_1$ and $\vec p_2$ found by using the previous recursive formula. The others can easily be found by using the symmetry under cyclic permutations of the momenta.
\begin{itemize}
\item
Integral $\mathcal{I}_{0, p_1 p_2} $:
\begin{equation}
 \mathcal{I}_{0, p_1 p_2} =
\int_k  \frac{1}{\vec k^2 | \vec k - \vec p_1|^2 | \vec k + \vec p_2|^2}  = i_{3,0,\{1,1,1\}}=\frac{1}{8 p_1 p_2 p_3}.
\end{equation}
\item
Integral $\mathcal{I}_{1, p_1 p_2}^{i_1   }$:
\be
 \mathcal{I}_{1, p_1 p_2}^{i_1   } =
\int_k  \frac{ k^{i_1} }{\vec k^2 | \vec k - \vec p_1|^2 | \vec k + \vec p_2|^2} =
2 p_1 ^{i_1} i_{3,1,\{1,2,1\}} - 2 p_2 ^{i_1} i_{3,1,\{2,1,1\}}.
\ee
\item
 Integral $\mathcal{I}_{2, p_1 p_2}^{i_1 i_2  } $:
\be
\mathcal{I}_{2, p_1 p_2}^{i_1 i_2  } =
\int_k \frac{ k^{i_1}k^{i_2} }{\vec k^2 | \vec k - \vec p_1|^2 | \vec k + \vec p_2|^2} =
\delta^{i_1 i_2} i_{3,1,\{1,1,1\}}+
4 p_1 ^{i_1} p_1^{i_2} i_{3,2,\{ 1,3,1\}}
\\
-4 p_1 ^{i_1} p_2^{i_2} i_{3,2,\{ 2,2,1\}}
-4 p_2 ^{i_1} p_1^{i_2} i_{3,2,\{ 2,2,1\}}
+4 p_2 ^{i_1} p_2^{i_2} i_{3,2,\{ 3,1,1\}}.
\ee
\end{itemize}
The integrals ${\I_3},\ {\I_4},\ {\I_5},\ {\I_6}$ can be found in the same way and contain $14$, $41$, $122$, and $365$ terms respectively. Furthermore, $ {\I_4},\ {\I_5},\ {\I_6}$ contain divergent terms which need to be regularised. The procedure is outlined in Appendix\hspace{0.1cm}\ref{app:reg}. The convergent ones were also checked  numerically.

\section{Regularisation}\label{app:reg}
In many of the computations performed, one arrives to expressions involving integrals of Bessel-K functions which could be divergent. In the following we provide the regularisation procedure focusing on the triple-K case since the simpler double-K strictly follows from analogous considerations. In the later subsections, we are going to focus on each case separately.

In the computation of the correlators we encounter triple-K integrals of the form
\be
I_{\alpha\{\beta_1, \beta_2, \beta_3\}} (p_1,p_2,p_3)= \int_0^\infty \d x \: x^{\alpha } \prod_{j=1}^3 p_j^{\beta_j} K_{\beta_j}(p_j x) .
\ee
The integral depends only on four parameters $(\alpha, \beta_j)$, since $p_j$ are the external momenta. The integral is always convergent at large values of $x$ due to the properties of the Bessel functions, while a divergence can come from the singularity at the lower limit  $x=0$. In particular, the integral converges if the following condition holds
\be
\alpha > \sum_{j=1}^{3} | \beta_j | -1.
\ee
Recall that we use $p_j$ to denote the modulus of the vectors $\vec p_j$ and therefore is always positive by construction.
In case the condition is not satisfied by the parameters, the integral is divergent and needs to be regularised. The regularisation procedure was laid down in Ref.~\cite{Bzowski:2013sza} and it is based on analytic continuation. We sum up the main steps here. One can introduce two additional real parameters $u$ and $v$ such that
\be
I_{\alpha\{\beta_1, \beta_2, \beta_3\}} \rightarrow  I_{\alpha+ u \epsilon \{\beta_1+ v \epsilon, \beta_2 + v \epsilon , \beta_3 + v \epsilon \}}.
\ee
The original integral is recovered in the limit of $\epsilon \rightarrow 0$. The limit exists and the analytically continued integral is independent of the choice of $u$ and $v$ except when one of the following conditions is met
\be \label{polecondition}
\alpha+1 \pm \beta_1 \pm \beta_2 \pm \beta_3 =-2n,
\ee
where $n$ is a non-negative integer. In these cases the solution contains a pole of the form $1/\epsilon$ and a dependence on $u$ and $v$ is still present. 
For example, 
one can perform the dimensional regularisation using  $u=v=-1/2$ since
\be
d \rightarrow d+2 u \epsilon, \qquad \Delta_j \rightarrow \Delta_j + (u+v) \epsilon.
\ee

\subsection{Double-K integrals}
The general solution of the double-K integral is
\begin{align} \label{e:I2K}
&I_{\alpha\{\beta_1, \beta_2\}} (p,p)= p^{\beta_1+\beta_2} \int_0^\infty \d  x \: x^{\alpha } K_{\beta_1}(p x) K_{\beta_2}(p x)
\nn
= \frac{2^{\alpha - 2}}{\Gamma(\alpha+1) p^{\alpha+1-\beta_1-\beta_2}} 
\\ 
&\times \Gamma \left( \frac{\alpha + \beta_1 + \beta_2+1}{2} \right) 
\Gamma \left( \frac{\alpha + \beta_1 - \beta_2+1}{2} \right) 
\Gamma \left( \frac{\alpha - \beta_1 + \beta_2+1}{2} \right) 
\Gamma \left( \frac{\alpha - \beta_1 - \beta_2+1}{2} \right),
\end{align}
valid for
\begin{equation}
\re( \alpha+1) > | \re \beta_1 | + | \re \beta_2 |.
\end{equation}
Thanks to the regularisation procedure
\be
I_{\alpha\{\beta_1, \beta_2\}} \rightarrow  I_{\alpha+ u \epsilon \{\beta_1+ v \epsilon, \beta_2 + v \epsilon \}},
\ee
 the preceding formula can also be used in the divergent cases and the result can be analytically continued by performing the limit $\epsilon \rightarrow 0$, after the eventual pole in $\epsilon$ is isolated.
For example
\be
I_{4,\{ 4,4\}}(p,p)=\frac{10395 \pi ^2}{512}p^{3} .
\ee

\subsection{Triple-K integrals}
The general solution of the triple-K integral is known to be
 \begin{align}
I_{\alpha\{\beta_1, \beta_2, \beta_3\}} (p_1,p_2,p_3)&= p_1^{\beta_1} p_2^{\beta_2} p_3^{\beta_3} \int_0^\infty {\rm d} x \: x^{\alpha} K_{\beta_1}(p_1 x) K_{\beta_2}(p_2 x) K_{\beta_3}(p_3 x) 
\nn \\
&= p_1^{\beta_1} p_2^{\beta_2} p_3^{\beta_3}
 \frac{2^{\alpha - 3}}{p_3^{\alpha+1}} \left[ A(\beta_1, \beta_2) + A(\beta_1, -\beta_2) + A(-\beta_1, \beta_2) + A(-\beta_1, -\beta_2) \right], \label{e:KKK}
\end{align}
where
\begin{align}
A(\beta_1, \beta_2) & = \left( \frac{p_1}{p_3} \right)^{\beta_1} \left( \frac{p_2}{p_3} \right)^{\beta_2} \Gamma \left( \frac{\alpha + \beta_1 + \beta_2 - \beta_3+1}{2} \right) \Gamma \left( \frac{\alpha + \beta_1 + \beta_2 + \beta_3+1}{2} \right) \Gamma(-\beta_1) \Gamma(-\beta_2)  \nn\\
&\times {F}_4 \left( \frac{\alpha + \beta_1 + \beta_2 - \beta_3+1}{2}, \frac{\alpha + \beta_1 + \beta_2 + \beta_3+1}{2}; \beta_1 + 1, \beta_2 + 1; \frac{p_1^2}{p_3^2}, \frac{p_2^2}{p_3^2} \right),
\end{align}
valid for
\begin{equation}\label{tkc}
\re (\alpha+1)> | \re \beta_1 | + | \re \beta_2 | + | \re \beta_3 |.
\end{equation}
The analytical continuation allows to use the preceding formula in the cases where the condition in Eq.~\eqref{tkc} is not met. Once the eventual pole in $\epsilon$ is  factorised, the limit $\epsilon \rightarrow 0$ can be found.
Some examples of regularised integrals appearing in our computations are the following (see Refs.~\cite{Bzowski:2013sza,Bzowski:2018fql})
\begin{align}
 I_{\frac{5}{2} + \epsilon \{\frac{3}{2}\frac{3}{2}\frac{3}{2}\}} &= - \left( \frac{\pi}{2} \right)^{3/2} \frac{1}{(p_1 + p_2 + p_3)^2} \left[ 2 p_1 p_2 p_3 + p_1^3 + p_2^3 + p_3^3 \right.\nn\\
& \qquad\qquad \left. +\:2 (p_1^2 p_2 + p_1 p_2^2 + p_1^2 p_3 + p_1 p_3^2 + p_2^2 p_3 + p_2 p_3^2) \right] + {\cal O}(\epsilon), \\
 I_{\frac{1}{2} + \epsilon \{\frac{3}{2}\frac{3}{2}\frac{3}{2}\}} &= \frac{1}{3} \left( \frac{\pi}{2} \right)^{3/2} \bigg[\frac{p_1^3 + p_2^3 + p_3^3}{\epsilon} - p_1p_2p_3 + (p_1^2p_2 + p_2^2p_1 + p_1^2p_3 + p_3^2p_1 + p_2^2p_3 + p_3^2p_2)  \nonumber \\
&  - (p_1^3 + p_2^3 + p_3^3){\rm ln}(p_1 + p_2 + p_3) + \frac{4}{3}(p_1^3 + p_2^3 + p_3^3)\bigg] .\label{e:exj0000}
\end{align}
Notice that the previous formulas were obtained with the choice $u=1$ and $v=0$.
Using the relations in Eq.~\eqref{iterative div} one can derive the  integrals needed in our computation.
For example
\begin{align}
I_{\frac{7}{2} \{\frac{3}{2}\frac{3}{2}\frac{5}{2}\}} &=  \bigg(-p_3 \frac{\partial}{\partial p_3} I_{\frac{5}{2} + \epsilon \{\frac{3}{2}\frac{3}{2}\frac{3}{2}\}} + 2 \cdot \frac{3}{2} I_{\frac{5}{2} + \epsilon \{\frac{3}{2}\frac{3}{2}\frac{3}{2}\}}\bigg)\bigg|_{\epsilon \rightarrow 0},
\\
I_{\frac{7}{2} \{\frac{5}{2}\frac{5}{2}\frac{5}{2}\}} &= \prod_{i=1}^{3} \bigg(-p_i \frac{\partial}{\partial p_i} + 3\bigg) I_{\frac{1}{2} + \epsilon \{\frac{3}{2}\frac{3}{2}\frac{3}{2}\}}\bigg|_{\epsilon \rightarrow 0}.\label{7/2 5/2 5/2 5/2}
\end{align}
Notice that the integral in Eq.~\eqref{7/2 5/2 5/2 5/2}, by having $\alpha=7/2$ and $\beta_i = 5/2$, does not satisfy the condition in Eq.~\eqref{polecondition}, thus no pole in $\epsilon$ is present and the analytically continued result is finite.

\section{Polarisation structures in the $X$ and $P$ basis}\label{pol}
In the following we provide the explicit formula for the structures $\cal E$ that appear in the shape of the graviton three-point function, see Eq.~\eqref{Est}. Here we choose  the $X$ and $P$ basis.
Assuming the external momenta $\vec p_i$, satisfying the momentum conservation, lying in the $(x,y)$-plane of a suitable reference frame, with $\vec p_1$ parallel to $\hat x$, one can define the polarisation tensors as
\begin{equation}
\epsilon^{ P}_{ij}\left(\varphi _i \right)=\frac{1}{\sqrt{2}}\left( {\begin{array}{ccc}
	-\sin^2\varphi_i\,\,\, & \cos\varphi_i \sin\varphi_i \,\,\, & 0\\
	\cos\varphi_i \sin\varphi_i \,\,\,&  -\cos^2\varphi_i \,\,\,& 0 \\
	0 \,\,\,& 0 \,\,\,& 1\\
	\end{array} } \right)\,,
\end{equation}
\begin{equation}
\epsilon^{X}_{ij}\left(\varphi _i \right)=\frac{1}{\sqrt{2}}\left( {\begin{array}{ccc}
0 \,\,\, & 0 \,\,\, & \sin\varphi_i\\
	0 \,\,\,&  0 \,\,\,& -\cos\varphi_i \\
	\sin\varphi_i \,\,\,& -\cos\varphi_i \,\,\,& 0\\
	\end{array} } \right)\,,
\end{equation}
where $\varphi_1 = 0$ and $\varphi_2$ and $\varphi_3$ identifies the angles between $\vec p_1$, $\vec p_2$ and $\vec p_1$, $\vec p_3$, respectively. The expressions appearing in Eq.~\eqref{Est} in the $\{P, X\}$ basis are then given by
\begin{eqnarray}
{\cal E}_1^{P}(\vp_1| \vp_2,\vp_2)&=& \frac{(p_1-p_2-p_3)(p_1+p_2-p_3)(p_1-p_2+p_3)(p_1+p_2+p_3)}{4\sqrt{2}p_1^2}, \nonumber \\
{\cal E}_1^{P}(\vp_2| \vp_3,\vp_3)&=&\frac{(-p_1+p_2-p_3)(p_1+p_2-p_3)(-p_1+p_2+p_3)(p_1+p_2+p_3)}{4\sqrt{2}p_2^2}, \nonumber \\
{\cal E}_1^{P}(\vp_3| \vp_1,\vp_1)&=& \frac{(p_1-p_2-p_3)(p_1+p_2-p_3)(p_1-p_2+p_3)(p_1+p_2+p_3)}{4\sqrt{2}p_3^2} ,
\end{eqnarray}
\begin{eqnarray}
{\cal E}_2^{PP}(\vp_1,\vp_2| \vp_2,\vp_3)&=& \frac{(p_1-p_2-p_3)(p_1+p_2-p_3)(p_1-p_2+p_3)(p_1+p_2+p_3)(p_1^2+p_2^2-p_3^2)}{16p_1^2p_2^2}, \nonumber \\
{\cal E}_2^{XX}(\vp_1,\vp_2| \vp_2,\vp_3)&=& -\frac{(p_1-p_2-p_3)(p_1+p_2-p_3)(p_1-p_2+p_3)(p_1+p_2+p_3)}{8p_1p_2}, \nonumber \\
{\cal E}_2^{PP}(\vp_2,\vp_3| \vp_3,\vp_1)&=&\frac{(-p_1+p_2-p_3)(p_1+p_2-p_3)(-p_1+p_2+p_3)(p_1+p_2+p_3)(-p_1^2+p_2^2+p_3^2)}{16p_2^2p_3^2}, \nonumber \\
{\cal E}_2^{XX}(\vp_2,\vp_3| \vp_3,\vp_1)&=&-\frac{(-p_1+p_2-p_3)(p_1+p_2-p_3)(-p_1+p_2+p_3)(p_1+p_2+p_3)}{8p_2p_3}, \nonumber \\
{\cal E}_2^{PP}(\vp_1,\vp_3| \vp_2,\vp_1)&=&\frac{(p_1-p_2-p_3)(p_1+p_2-p_3)(p_1-p_2+p_3)(p_1+p_2+p_3)(p_1^2-p_2^2+p_3^2)}{16p_1^2p_3^2}, \nonumber \\
{\cal E}_2^{XX}(\vp_1,\vp_3| \vp_2,\vp_1)&=&-\frac{(p_1-p_2-p_3)(p_1+p_2-p_3)(p_1-p_2+p_3)(p_1+p_2+p_3)}{8p_1p_3}, 
\end{eqnarray}
\begin{eqnarray}
{\cal E}_4^{PPP}(\vp_1,\vp_2,\vp_3|\vp_4,\vp_5)&=& \frac{-(p_1^2-p_2^2)^4+2(p_1^2-p_2^2)^2(p_1^2+p_2^2)p_3^2-2(p_1^2+p_2^2)p_3^6+p_3^8}{32\sqrt{2}p_1^2p_2^2p_3^2} ,\nonumber \\
{\cal E}_4^{XXP}(\vp_1,\vp_2,\vp_3|\vp_4,\vp_5)&=& -\frac{(p_1-p_2-p_3)(p_1+p_2-p_3)(p_1-p_2+p_3)(p_1+p_2+p_3)}{8\sqrt{2}p_1p_2},\nonumber \\
{\cal E}_4^{XPX}(\vp_1,\vp_2,\vp_3|\vp_4,\vp_5)&=& \frac{(p_1-p_2-p_3)(p_1+p_2-p_3)(p_1-p_2+p_3)(p_1+p_2+p_3)(p_1^2-p_2^2-p_3^2)}{16\sqrt{2}p_1p_2^2p_3} ,
\nonumber \\
{\cal E}_4^{PXX}(\vp_1,\vp_2,\vp_3|\vp_4,\vp_5)&=& -\frac{(p_1-p_2-p_3)(p_1+p_2-p_3)(p_1-p_2+p_3)(p_1+p_2+p_3)(p_1^2-p_2^2+p_3^2)}{16\sqrt{2}p_1^2p_2p_3} ,\nonumber
\end{eqnarray}
\begin{eqnarray}
{\cal E}_4^{PPP}(\vp_3,\vp_2,\vp_1|\vp_1,\vp_3) &=& \frac{(p_1-p_2-p_3)(p_1+p_2-p_3)(p_1-p_2+p_3)(p_1+p_2+p_3)}{32\sqrt{2}p_1^2p_2^2p_3^2}  \nonumber \\
&\times &(p_1^2+p_2^2-p_3^2)(p_1^2-p_2^2+p_3^2), \nonumber \\
{\cal E}_4^{XXP}(\vp_3,\vp_2,\vp_1|\vp_1,\vp_3) &=& -\frac{(-p_1+p_2-p_3)(p_1+p_2-p_3)(-p_1+p_2+p_3)(p_1+p_2+p_3)}{8\sqrt{2}p_2p_3} ,\nonumber \\
{\cal E}_4^{XPX}(\vp_3,\vp_2,\vp_1|\vp_1,\vp_3) &=& -\frac{(p_1-p_2-p_3)(p_1+p_2-p_3)(p_1-p_2+p_3)(p_1+p_2+p_3)(p_1^2+p_2^2-p_3^2)}{16\sqrt{2}p_1p_2^2p_3} ,\nonumber \\
{\cal E}_4^{PXX}(\vp_3,\vp_2,\vp_1|\vp_1,\vp_3)&=& -\frac{(p_1-p_2-p_3)(p_1+p_2-p_3)(p_1-p_2+p_3)(p_1+p_2+p_3)(p_1^2-p_2^2+p_3^2)}{16\sqrt{2}p_1p_2p_3^2} ,\nonumber \\ 
{\cal E}_4^{PPP}(\vp_1,\vp_3,\vp_2|\vp_2,\vp_1)&=&- \frac{(p_1-p_2-p_3)(p_1+p_2-p_3)(p_1-p_2+p_3)(p_1+p_2+p_3)(p_1^2-p_2^2-p_3^2)}{32\sqrt{2}p_1^2p_2^2p_3^2}\nonumber\\
&\times &(p_1^2+p_2^2-p_3^2), \nonumber \\
{\cal E}_4^{XXP}(\vp_1,\vp_3,\vp_2|\vp_2,\vp_1) &=& -\frac{(p_1-p_2-p_3)(p_1+p_2-p_3)(p_1-p_2+p_3)(p_1+p_2+p_3)}{8\sqrt{2}p_1p_3} ,\nonumber \\
{\cal E}_4^{XPX}(\vp_1,\vp_3,\vp_2|\vp_2,\vp_1) &=& \frac{(p_1-p_2-p_3)(p_1+p_2-p_3)(p_1-p_2+p_3)(p_1+p_2+p_3)(p_1^2-p_2^2-p_3^2)}{16\sqrt{2}p_1p_2p_3^2} ,\nonumber \\
{\cal E}_4^{PXX}(\vp_1,\vp_3,\vp_2|\vp_2,\vp_1) &=& -\frac{(p_1-p_2-p_3)(p_1+p_2-p_3)(p_1-p_2+p_3)(p_1+p_2+p_3)}{16\sqrt{2}p_1^2p_2p_3}
\nonumber\\
&\times & (p_1^2+p_2^2-p_3^2),  
\end{eqnarray}
\begin{eqnarray}
{\cal E}_3^{PP}(\vp_1,\vp_2)&=& \frac{1}{8}\bigg(4+\frac{(p_1^2+p_2^2-p_3^2)^2}{p_1^2p_2^2}\bigg) , \nonumber \\
{\cal E}_3^{XX}(\vp_1,\vp_2)&=& -\frac{p_1^2+p_2^2-p_3^2}{2p_1p_2} , \nonumber \\
{\cal E}_3^{PP}(\vp_2,\vp_3)&=&  \frac{p_1^4+p_2^4+6p_2^2p_3^2+p_3^4-2p_1^2(p_2^2+p_3^2)}{8p_2^2p_3^2}, \nonumber \\
{\cal E}_3^{XX}(\vp_2,\vp_3)&=& -\frac{-p_1^2+p_2^2+p_3^2}{2p_2p_3}, \nonumber \\
{\cal E}_3^{PP}(\vp_1,\vp_3)&=& \frac{1}{8}\bigg(4+\frac{(p_1^2-p_2^2+p_3^2)^2}{p_1^2p_3^2}\bigg) , \nonumber \\
{\cal E}_3^{XX}(\vp_1,\vp_3)&=& -\frac{p_1^2-p_2^2+p_3^2}{2p_1p_3},
\end{eqnarray}
\begin{eqnarray}
{\cal E}_5^{PPP}(\vp_1,\vp_2,\vp_3)&=& -\frac{(p_1-p_2-p_3)(p_1+p_2-p_3)(p_1-p_2+p_3)(p_1+p_2+p_3)(p_1^2+p_2^2+p_3^2)}{16\sqrt{2}p_1^2p_2^2p_3^2}, \nonumber\\
{\cal E}_5^{XXP}(\vp_1,\vp_2,\vp_3)&=& \frac{(p_1-p_2-p_3)(p_1+p_2-p_3)(p_1-p_2+p_3)(p_1+p_2+p_3)}{8\sqrt{2}p_1p_2p_3^2}, \nonumber\\
{\cal E}_5^{XPX}(\vp_1,\vp_2,\vp_3)&=& \frac{(p_1-p_2-p_3)(p_1+p_2-p_3)(p_1-p_2+p_3)(p_1+p_2+p_3)}{8\sqrt{2}p_1p_2^2p_3}, \nonumber\\
{\cal E}_5^{PXX}(\vp_1,\vp_2,\vp_3)&=& \frac{(p_1-p_2-p_3)(p_1+p_2-p_3)(p_1-p_2+p_3)(p_1+p_2+p_3)}{8\sqrt{2}p_1^2p_2p_3},
\end{eqnarray}
and the other  combinations are vanishing.

\section{Polarisation structures in the chiral basis}\label{epsi}
In this Appendix we provide the explicit formula for the structures $\cal E$ in Eq.~\eqref{Est} in the chiral basis. With respect to the $X$ and $P$ basis one has
\be
\epsilon^{\rm R}_{ij}=\frac{\epsilon^{P}_{ij}+i \epsilon^{X}_{ij}}{\sqrt{2}}\,\,\,\,{\rm and}\,\,\,\,\epsilon^{\rm L}_{ij}=\frac{\epsilon^{P}_{ij}-i \epsilon^{X}_{ij}}{\sqrt{2}}.
\ee
We can express the structures $\cal E$ as a function of the ratios between the external momenta, i.e $r_2=\frac{p_2}{p_1}$ and $r_3=\frac{p_3}{p_1}$, and for generic polarisation $\lambda_1,\lambda_2,\lambda_3$ of the chiral basis, that is  $\lambda=\{\rm R,L\}$. From the explicit formula provided in the previous appendix, one can write the polarisation tensors in the new basis as
\begin{equation}
\epsilon^{\lambda}_{ij}\left(\varphi _i \right)=\frac{1}{2}\left( {\begin{array}{ccc}
   -\sin^2\varphi_i\,\,\, & \cos\varphi_i \sin\varphi_i \,\,\, &  i \lambda \sin\varphi_i\\
   \cos\varphi_i \sin\varphi_i \,\,\,&  -\cos^2\varphi_i \,\,\,& - i  \lambda  \cos\varphi_i \\
     i  \lambda \sin\varphi_i \,\,\,&  - i  \lambda \cos\varphi_i \,\,\,& 1\\
  \end{array} } \right),
\end{equation}
where again $\varphi_1 = 0$ and $\varphi_2$ and $\varphi_3$ identifies the angles between $\vec p_1$, $\vec p_2$ and $\vec p_1$, $\vec p_3$, respectively. The ${\rm R}$ polarisation corresponds to $\lambda=1$ while the ${\rm L}$ polarisation corresponds to $\lambda=-1$. One can check that this construction satisfies the normalisation condition in Eq.~\eqref{normepsilon} and it is  also transverse and traceless.
The expressions appearing in Eq.~\eqref{Est} in the \{\rm R, L\} basis are 
\begin{eqnarray}
	{\cal E}_1^{\lambda_1}(\vp_1| \vp_2,\vp_2)&=& \frac{p_1^2}{8} \left[ r_2^4+(-1+r_3^2)^2 -2r_2^2( 1+r_3^2)\right], \nonumber \\
	{\cal E}_1^{\lambda_1}(\vp_2| \vp_3,\vp_3)&=& \frac{p_1^2}{8r_2^2} \left[ r_2^4+(-1+r_3^2)^2 -2r_2^2( 1+r_3^2)\right] , \nonumber \\
	{\cal E}_1^{\lambda_1}(\vp_3| \vp_1,\vp_1)&=& \frac{p_1^2}{8r_3^2} \left[ r_2^4+(-1+r_3^2)^2 -2r_2^2( 1+r_3^2)\right],
\end{eqnarray}
\begin{eqnarray}
	{\cal E}_2^{\lambda_1\lambda_2}(\vp_1,\vp_2| \vp_2,\vp_3)&=&\frac{p_1^2}{32r_2^2}\left[r_2^4+(-1+r_3^2)^2-2r_2^2(1+r_3^2)\right]\left[1+r_2^2-r_3^2+2r_2\lambda_1\lambda_2\right] , \nonumber \\
	{\cal E}_2^{\lambda_2\lambda_3}(\vp_2,\vp_3| \vp_3,\vp_1)&=&\frac{p_1^2}{32r_2^2r_3^2}\left[r_2^4+(-1+r_3^2)^2-2r_2^2(1+r_3^2)\right]\left[-1+r_2^2+r_3^2+2r_2r_3\lambda_2\lambda_3\right], \nonumber \\
	{\cal E}_2^{\lambda_1\lambda_3}(\vp_1,\vp_3| \vp_2,\vp_1)&=&\frac{p_1^2}{32r_3^2}\left[r_2^4+(-1+r_3^2)^2-2r_2^2(1+r_3^2)\right]\left[1-r_2^2+r_3^2+2r_3\lambda_1\lambda_3\right],
\end{eqnarray}
\begin{eqnarray}
	{\cal E}_3^{\lambda_1\lambda_2}(\vp_1,\vp_2)&=& \frac{r_2^4 -2r_2^2(-3+r_3^2)+(-1+r_3^2)^2+4r_2^3\lambda_1\lambda_2-4r_2(-1+r_3^2)\lambda_1\lambda_2}{16r_2^2} , \nonumber \\
	{\cal E}_3^{\lambda_2\lambda_3}(\vp_2,\vp_3)&=& \frac{1-2r_2^2+r_2^4-2r_3^2+6r_2^2r_3^2+r_3^4+4r_2r_3(-1+r_2^2+r_3^2)\lambda_2\lambda_3}{16r_2^2r_3^2} , \nonumber \\
	{\cal E}_3^{\lambda_1\lambda_3}(\vp_1,\vp_3)&=& \frac{1+r_2^4+6r_3^2+r_3^4+4(r_3+r_3^3)\lambda_1\lambda_3-2r_2^2(1+r_3^2+2r_3\lambda_1\lambda_3)}{16r_3^2},
\end{eqnarray}
\begin{eqnarray}
	&&{\cal E}_4^{\lambda_1\lambda_2\lambda_3}(\vp_1,\vp_2,\vp_3|\vp_4,\vp_5)= -\frac{p_1^2}{128r_2^2r_3^2} \left[(-1+r_2^2)^2-2(1+r_2^2)r_3^2+r_3^4\right] \nonumber \\
	&&\times \left[1-2r_2^2+r_2^4-r_3^4-4r_2r_3^2\lambda_1\lambda_2-2r_3\left((-1+r_2^2+r_3^2)\lambda_1+r_2(1-r_2^2+r_3^2)\lambda_2\right)\lambda_3 \right], \nonumber \\
	&& {\cal E}_4^{\lambda_3\lambda_2\lambda_1}(\vp_3,\vp_2,\vp_1|\vp_1,\vp_3) = -\frac{p_1^2}{128r_2r_3^2}\left[(-1+r_2^2)^2-2(1+r_2^2)r_3^2+r_3^4\right] \nonumber \\
	&&\times 
	\left[(-1+r_2^2-r_3^2)(1+r_2^2-r_3^2+2r_2\lambda_1\lambda_2)-2r_3\left((1+r_2^2-r_3^2)\lambda_1+2r_2\lambda_2\right)\lambda_3 \right], \nonumber \\
	&& {\cal E}_4^{\lambda_1\lambda_3\lambda_2}(\vp_1,\vp_3,\vp_2|\vp_2,\vp_1) = 
	\frac{p_1^2}{128r_2^2r_3^2}\left[(-1+r_2^2)^2-2(1+r_2^2)r_3^2+r_3^4\right] \nonumber \\
	&&\times 
	\left[(-1+r_2^2+r_3^2)(1+r_2^2-r_3^2+2r_2\lambda_1\lambda_2)+2r_2r_3(2r_2\lambda_1+\lambda_2+r_2^2\lambda_2-r_3^2\lambda_2)\lambda_3\right],
\end{eqnarray}
\begin{eqnarray}
	{\cal E}_5^{\lambda_1\lambda_2\lambda_3}(\vp_1,\vp_2,\vp_3)&=&-\frac{1}{64 r_2^2 r_3^2}\left[2 r_2+(1+r_2^2-r_3^2)\lambda_1\lambda_2\right]
	\left[-2 r_3+(-1+r_2^2-r_3^2)\lambda_1\lambda_3\right]\nonumber\\
	&\times&\left[2 r_2 r_3+(-1+r_2^2+r_3^2)\lambda_2\lambda_3\right].
\end{eqnarray}

\end{document}